%% file: paper.tex
\documentclass[aps,amssymb,amsmath,prl,preprint,floatfix]{revtex4}
\usepackage{graphicx}
\usepackage{color}

\def\d {{\rm d}}
\def\mcalu{{\mathring{\cal U}}}

\def\tensor#1{{\underline{\underline {#1}}}}
\def\ttensor#1{{\underline{\underline{\underline {#1}}}}}
\def\vec#1{{{\underline {#1}}}}

\def\bn{\begin{itemize}}
\def\en{\end{itemize}}
\def\Tr{\text{Tr}}
\def\xigamma{{\bf\vec\Xi}}

\begin{document}
\bibliographystyle{prsty}

\title{Sum rules for the quasi-static and visco-elastic response of disordered solids at zero temperature}
\author{Ana\"el Lema\^{\i}tre$^{(1,3)}$}
\altaffiliation[Current address:]{Institut Navier -- LMSGC, 2 all\'ee K\'epler, 77420 Champs-sur-Marne}
\author{Craig Maloney$^{(1,2)}$}
\affiliation{
$^{(1)}$ Department of physics, University of California, Santa Barbara, California 93106, U.S.A.}
\affiliation{$^{(2)}$ Lawrence Livermore National Lab - CMS/MSTD, Livermore, California 94550, U.S.A.}
\affiliation{$^{(3)}$ L.M.D.H. - Universite Paris VI, UMR 7603,
4 place Jussieu, 75005  Paris - France}
\date{\today}

\begin{abstract}
We study exact results concerning the non-affine displacement fields 
observed by Tanguy {\it et al\/} [Europhys. Lett. {\bf 57}, 423  (2002),
Phys. Rev. B {\bf 66},  174205  (2002)] and 
their contributions to elasticity.
A normal mode analysis permits us to estimate the dominant contributions
to the non-affine corrections to elasticity and relate these corrections 
to the correlator of a fluctuating force field.
We extend this analysis to 
the visco-elastic dynamical response of the system.
\end{abstract}
\maketitle

A straightforward estimate of the elastic constants of simple crystals
can be performed in the (classical) zero temperature limit:
the relative initial positions of atoms are known;
elementary deformations are homogeneous even at the microscopic level.
It is thus a simple task to add up all contributing interactions.
These assumptions--zero temperature and homogeneous displacement 
of the particles--constitute the basis of the Born-Huang 
theory.~\cite{huang50,BH54}
These assumptions can also be used to estimate the elastic constants 
of a disordered structure: they provide approximate expressions involving 
integrals over the pair correlation; in liquid theory,
these expressions correspond to the infinite frequency moduli.~\cite{Zwanzig65,HM86}
Of course, the two assumptions of zero temperature and homogeneous displacement
are not valid in general and corrections to the Born-Huang approximation are expected 
to arise from the failure of either.

Early studies by Squire, Holt and Hoover,~\cite{SHH68,HHS69}
focused on thermal contributions to elasticity in crystals.
More recently, a surge of interest for athermal materials,
like granular materials or foams, attracted some attention to 
corrections to the Born-Huang approximation which arise
solely from the non-trivial structure of the potential 
energy landscape.~\cite{LGL+96,LL97,RadjaiR02,WTB+02,TWL+02}
Namely, in disordered solids at zero temperature,
the assumption that particles follow homogeneous (affine)
displacement fields is incorrect: when a material is strained--even by vanishingly 
small amounts of deformation--particles minimize the potential 
energy of the system by following non-affine displacements.
This idea was recently recognized in numerical simulations of compressed emulsions~\cite{LGL+96} and Lennard-Jones glasses.~\cite{WTB+02,TWL+02}
In particular, Tanguy  {\it et al}~\cite{WTB+02,TWL+02} have clearly shown that non-affine corrections to the Born-Huang approximation hold in the continuum limit and  amount to an important fraction of the Born-Huang term itself. It is thus essential to understand these corrections well if we ever want to be able to construct approximations to the elastic constants of amorphous materials.

Formal expressions for the 
non-affine (or ``inhomogeneous'', in the language of Wallace~\cite{wallace72})
corrections to the Born-Huang approximation 
were written early on~\cite{wallace72,lutsko88,lutsko89,ZJ96,ZJ02}.
These formal expressions have been used almost exclusively as a tool 
to calculate elastic constants in computer simulations,
but were given little attention in light of their 
basic importance.
We believe that this arose from two limitations:
(i) prior works remained at an essentially technical level,
aiming merely to provide tools for numerical simulations
(ii) the derivations of formal analytical expressions for elastic constants
have always relied on simplifying assumptions--either dealing with 
an infinite system or valid at zero stress.
These simplifications make it difficult to ascertain the domain of 
validity of various formulae or symmetries.
In response to these issues, we wish to attack this problem 
from two opposite angles: provide an even more systematic and 
general treatment than before, yet relate this formalism to numerical
observations similar to Tanguy {\it et al\/}.~\cite{WTB+02,TWL+02}
We thus hope that our treatment could serve, 
at least, as an introduction to the subject.

 From the formalism, we wish to extract information about the contribution
of various scales to the non-affine corrections to elasticity: the question
we have at heart is whether these corrections originate from small or large
scales, or involve a broad distribution of scales as suggested by the 
observation of vortex-like patterns.~\cite{RadjaiR02} This question is directly related to 
the existence of a continuous limit for elastic properties 
of amorphous structures.~\cite{WTB+02,TWL+02}
To address these questions, we perform a normal mode decomposition of the non-affine 
displacement field: it permits quantifying the contribution 
of every frequency shell to the non-affine corrections to elastic constants. 
In the large size limit, these contributions seem to be self-averaging quantities (in the sense that an ensemble average over subsystems will produce results which are equivalent to a single large system).
The existence of this self-averaging property leads to an expression for 
the elastic constants in the continuum limit
which resembles the sum rules of liquid theory.

Finally, our attention was attracted by several related issues 
in the recent literature. 
Studies of sound propagation and attenuation in 
granular materials~\cite{liu92,JCV99,MGJ+04,Som05} 
or of the visco-elastic response of dense emulsions and foams,~\cite{MBW95,LGL+96,LRM+96,LL97,MLG+97}
emphasize the need for a deep theoretical understanding of
the visco-elastic response of disordered solids.
Related experiments by McKenna and coworkers indicated that features
of the visco-elastic response of amorphous materials are related to 
measurable changes in their elastic constants.~\cite{OM02}
We thus complement the study of static response by a study of dynamical response,
and derive a formal expression for the visco-elastic moduli.
We shall establish that the relaxation spectrum is directly and simply related 
to a correlator emerging from the normal mode analysis.

The present work is meant to present the general framework of our analysis,
which will be the basis of future numerical studies.
Although, here, we will use numerical simulations to illustrate 
analytical developments, the main core of our numerical study
will be presented in a dedicated article.~\cite{Inprep}

\section{1. Non-affine displacement fields}

In this work we consider the mechanical response of an amorphous solid quenched
at zero temperature. Our formalism permits dealing explicitly with finite size
systems: it rests on the idea that, during a quench at zero temperature, 
any finite size system relaxes toward one of many local minima in 
the potential energy landscape.~\cite{stillinger88} Being at zero temperature, 
the system is then prescribed to lie at this minimum at all times.
Small external perturbations are then expected to induce continuous changes
in the local minimum. Large external perturbations may induce the vanishing of the 
local minimum occupied by the system: this vanishing occurs when the basin 
of attraction 
of this minimum reduces to a single point, that is when the minimum collides with 
at least one saddle point.~\cite{ML97,ML98,ML99,ML04b} 

The difference between small, continuous changes of the local configuration and 
\begin{figure}
\includegraphics[width=\textwidth]{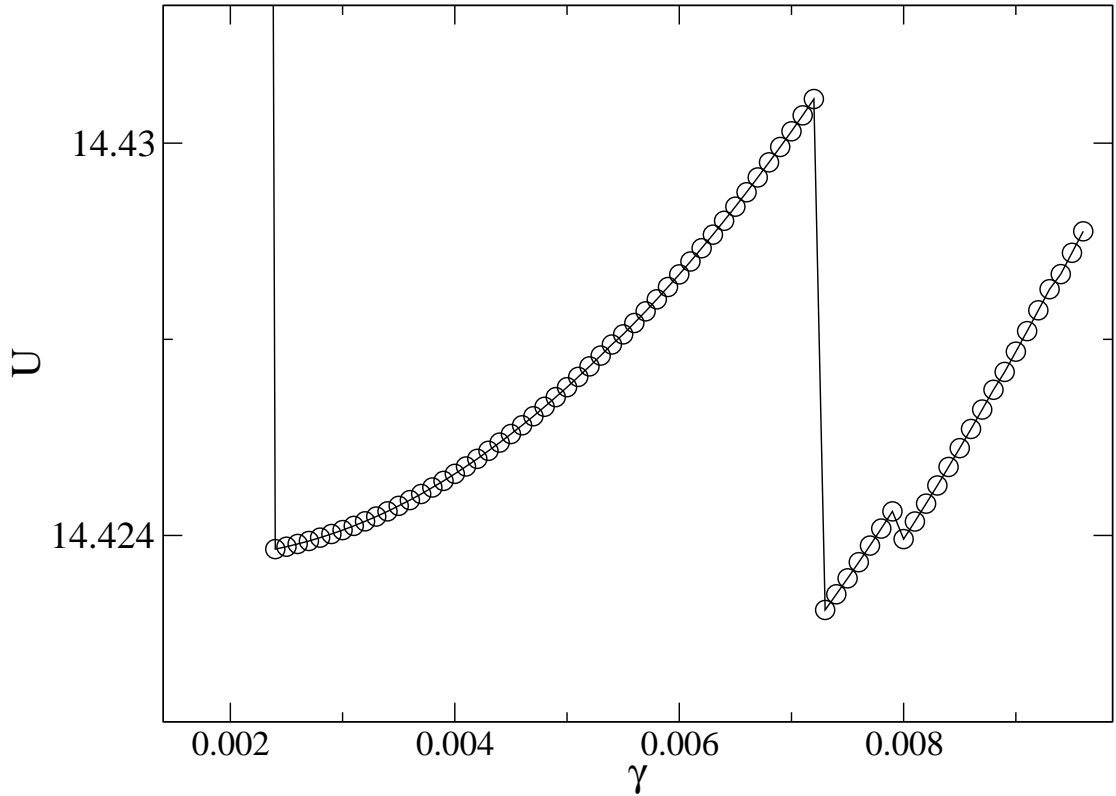}
\caption{A typical energy vs. strain curve obtained in an athermal quasistatic simulation as described in the text.  Note the smooth elastic segment which terminates at a plastic discontinuity. } 
\label{fig:strain}
\end{figure}
catastrophic events is exemplified figure~\ref{fig:strain} where the response of 
an amorphous system is shown as a function of shear. 
The parameter $\gamma$ measures the total shear deformation from a quenched state.  
In this picture, continuous segments are associated
with shear-induced changes of local minima and discontinuities to their vanishing. 
After a minimum has disappeared, 
the system, coupled to a zero temperature thermal bath, relaxes
in search of a new minimum in the potential energy landscape.~\cite{ML04,ML04b}

The separation of continuous changes of the local minimum 
and catastrophic events is, of course, not limited to shear deformation.
In a more general context, suppose that we denote by $\gamma$ the amplitude of 
any external drive applied to the system. Suppose that the system lies at a minimum 
which, for $\gamma=\gamma_0$, is far from any catastrophe.
Let us denote $\delta\gamma=\gamma-\gamma_0$ the amplitude 
of a perturbation of the external drive around $\gamma_0$. 
If we were to slowly increase the amplitude of the perturbation from 0, 
the system would smoothly follow a trajectory $\vec r_i(\gamma)$ 
in configuration space 
for all amplitudes $\delta\gamma<\epsilon$ such that the basin of the local minimum 
remains non-vanishing. 
If the Hessian is non-degenerate, this trajectory $\vec r_i(\delta\gamma)$ is unique.
If we now stop the perturbation at some $\delta\gamma_{\rm max}<\epsilon$ 
then revert the drive down to $\delta\gamma=0$, the system would simply follow 
the trajectory $\vec r_i(\gamma)$ from $\gamma=\gamma_0+\delta\gamma_{\rm max}$ to 
$\gamma_0$, backward. In this sense we will say that the continuous segments are 
microscopically reversible. 

When the system is driven quasi-statically along these continuous segments,
no energy is dissipated.
The reason is that at each point along these segments, the system is at 
mechanical equilibrium: the force applied to any particle is exactly zero and
zero forces do no work.
The quasi-static response corresponds to the thermodynamically reversible, elastic,
part of the mechanical response.
We will see, however, that energy is dissipated when the system is
driven at finite deformation rates along these continuous segments.
This dissipation results from the fact that finite deformation rates induce
non-zero forces which dissipate energy via the
coupling with the zero temperature thermal bath.

\subsection{1.1 Notation}
We shall let underline and double-underlines respectively 
indicate vectors and tensors 
referred to a fixed Cartesian system $(x_1, x_2, x_3)$.
We shall also use the convention that Greek indices refer to Cartesian
components of vectors or tensors, while Roman indices refer to the particle numbers.
Bold type denotes fields which are defined on every particle in the material:
$$
{\bf f} = \{f_i\}_{i\in\{1,\ldots, n\}}\quad,\quad\quad 
{\bf\tensor A} = \{\left(A_{i,\alpha\beta}\right)\}_{i\in\{1,\ldots, n\}}
\quad.
$$
Dots and double dots indicate matrix products and summation convention 
is always applied on repeated (Greek and Latin) indices.
By convention, we also write:
$$
\left(\frac{\partial {\vec A}}{\partial {\vec r}}\right)_{\alpha\beta}
=
\left(\frac{\partial {A_\alpha}}{\partial r_{\beta}}\right)
\quad\text{and}\quad
\left(\frac{\partial^2 {\vec A}}{\partial {\vec r}\partial {\vec s}}\right)_{\alpha\beta\gamma}
=
\left(\frac{\partial^2 {A_\alpha}}{\partial r_{\beta}\partial s_{\gamma}}\right)
\quad.
$$
The superscript $^T$ indicates the transpose of a matrix, $^{-1}$ its inverse,
and $^{-T}$ the inverse of its transpose. The identity matrix
is denoted $\tensor 1$, and its components $\delta_{\alpha\beta}$.

\subsection{1.2 Displacement fields}

In this work, utilizing the formalism underlying
the Andersen-Parrinello-Rahman
theory~\cite{andersen80,parrinello81,parrinello82,ray83,ray84,ray85}
we shall focus on the situation 
where a system of particles is contained in a periodic simulation cell.
The formalism we describe can easily be adapted to 
study the deformation of a material confined between walls:
to do so, it is sufficient to embed the whole system--confined particles plus walls--
in the cell and mandate that the particles constituting the walls
affinely follow its deformation.

The shape of the simulation cell
(which is, by construction, a parallelepiped)
 is represented by the set of $d=2$ or $3$ Bravais
vectors: $\tensor h =(\vec a,\vec b)$ or $\tensor h =(\vec a,\vec b,\vec c)$;
its volume is $V=\det(\tensor h)=|\tensor h|$.
We consider a system of $N$ particles, with positions ${\bf\vec r}=\{\vec r_i\}$ in real space.
The interaction potential ${\cal U}({\bf\vec r},\tensor h)$ depends on the positions
of the particles but also on the shape of the simulation cell
which enforces boundary conditions.

``Macroscopic'' deformations of the sample are performed by changing the Bravais vectors. 
Since we are concerned with variations of the local minimum around some 
reference configuration, we will often use a \emph{reference} configuration $\mathring{\tensor h}$ 
of the cell and compare it with a \emph{current} value $\tensor h$.
Following Ray and Rahman,~\cite{ray83,ray84,ray85} we introduce a transformation of 
particle coordinates which maps any vector ${\vec r}$
onto a cubic reference cell:
$$
{\vec r} = \tensor h.{\vec s}\quad,
$$
with $s_\alpha\in[-0.5,0.5]$.
If we change the cell coordinate from $\mathring{\tensor h}$ to $\tensor h$
and require that all particles affinely follow the deformation of the cell,
any particle at point $\mathring{\vec r}$ is mapped
onto ${\vec r}=\tensor h.\mathring{\tensor h}^{-1}.\mathring{\vec r}$.
We denote $\tensor F=\tensor h.\mathring{\tensor h}^{-1}$ which,
in the usual language of elasticity, is the deformation gradient tensor.~\cite{salencon01,slaughter02}
Once the reference frame $\mathring{\tensor h}$ is specified,
any configuration $(\{{\vec r}_i\},\tensor h)$ of the system 
can be parameterized by a pair 
$(\{\mathring{\vec r}_i\},\tensor F)$, with the convention that: 
${\vec r}_i=\tensor F.\mathring{\vec r}_i$.
In this parameterization, changes in $\tensor F$ correspond to affine transformations of 
all the particles following the cell shape, while
changes in $\mathring{\vec r}_i$
correspond to the non-affine part of the displacement of the particles.

At zero temperature, an infinitesimal deformation of the system is often performed
in two steps. First, starting from a local minimum $\{\mathring{\vec r}_i\}$ 
at $\mathring{\tensor h}$, the particle coordinates affinely follow the change
of the cell coordinate from $\mathring{\tensor h}$ to $\tensor h$.
The $\{\mathring{\vec r}_i\}$ remain constant.
The real-space position of particle $i$ is thus mapped from $\mathring{\vec r}_i$   
onto $\tensor F\,.\,\mathring{\vec r}_i$.
Second, the particles are allowed to relax to the nearest equilibrium position, 
$\tensor h$ being fixed. They reach new positions $\{{\vec r}_i\}$ which differ
in general from $\{\tensor F\,.\,\mathring{\vec r}_i\}$.
The non-affine part of the deformation is then characterized by the displacements as viewed in 
the reference frame, $\{\tensor F^{-1}.{\vec r}_i\}$.
For small displacements, the particles continuously follow changes of local minima.
The real space positions of the particles at equilibrium are thus a continuous function 
of $\tensor h$ (on some interval of strains), and we could denote these equilibria 
as $\{{\vec r}_i(\tensor h)\}$. Likewise, the continuous changes of local minima
are most readily studied by monitoring
changes in the \emph{reference} co-ordinates: 
$\{\mathring{\vec r}_i(\tensor h)=\tensor F^{-1}.{\vec r}_i(\tensor h)\}$.

\subsection{1.3 Equilibrium trajectories}

By definition, elastic constants are second order derivatives of the energy with respect
to strain.
Since strain is characterized by second order tensors, 
the elastic constants are fourth order tensors. Before presenting the details of 
the tensorial formalism, however, it seemed pedagogically 
sounder to us to first consider 
the simple situation where the 
shape of the cell can be parameterized by a single degree of freedom.

Suppose then that we prescribe the tensor $\tensor h(\gamma)$ as a function of 
a scalar parameter $\gamma$. 
For varying $\gamma$, so long as the local minimum does not vanish, the system follows a continuous
trajectory in configuration space as illustrated on figure~\ref{fig:strain}. 
Given a reference cell $\mathring{\tensor h} = \tensor h(\gamma_0)$ at $\gamma_0$,
the energy functional can be written either as a function of 
${\bf\vec r}$ and $\gamma$: ${\cal U}({\bf\vec r},\gamma)$;
or as a function of $\mathring{\bf\vec r}$ and $\gamma$:
$\mathring{\cal U}(\mathring{\bf\vec r},\gamma)\equiv {\cal U}(\tensor F(\gamma).\mathring{\bf\vec r},\gamma)$.
We introduce the notation $\mcalu$ to emphasize that--contrarily 
to ${\cal U}$--this function is defined after a choice of reference cell 
with Bravais matrix $\mathring{\tensor h}$. 
Changing $\gamma$ for fixed $\{\mathring{\vec r}_i\}$
corresponds to
performing an affine strain of the whole
system--the particles and the boundary.

When particles are constrained to follow deformation-induced changes of a local minimum,
their real-space positions $\{{\vec r}_i(\gamma)\}$ and 
the corresponding non-affine displacements $\{\mathring{\vec r}_i(\gamma)\}$ are now 
one-parameter functions of $\gamma$.
An equation of motion for the non-affine displacement fields derives from a straightforward
application of the implicit function theorem:
The trajectory is specified by the condition that the system is always at mechanical equilibrium:
\begin{equation}
\vec f_i=-\frac{\partial {\cal U}}{\partial \vec r_i}
=-\frac{\partial\mathring{\cal U}}{\partial\mathring{\vec r}_i}.{\tensor F}^{-1}=0
\quad.
\end{equation}
Note that the second equality is a property of the point derivatives of any observable
of the form, $A(\{{\vec r}_{i}=\tensor F.\mathring{\vec r}_{i}\})$:
\begin{equation}
\label{eqn:deriv:point}
\frac{\partial A}{\partial \mathring{\vec r}_i}=
\frac{\partial A}{\partial{\vec r}_i}\,.\,\tensor F
\quad,
\end{equation}
and is derived in appendix~A.
Differentiating the condition $\frac{\partial\mathring{\cal U}}{\partial\mathring{\vec r}_i}=0$ once
with respect to $\gamma$ leads to:
\begin{equation}
\label{eqn:motion:gamma}
\frac{\partial^2\mathring{\cal U}}{\partial\mathring{\vec r}_i\partial\mathring{\vec r}_j}.
\frac{{\cal D}\,\mathring{\vec r}_j}{{\cal D}\,\gamma}+
\frac{\partial^2\mathring{\cal U}}{\partial\mathring{\vec r}_i\partial\gamma}=0
\qquad,
\end{equation}
where the symbol ${\cal D}$ is introduced to indicate derivatives which are taken
under the constraints of mechanical equilibrium.
This equation is formally valid for all $\gamma\ne\gamma_0$,
even though it will primarily be used in the limit $\gamma\to\gamma_0$ (i.e.
${\tensor h}\to\mathring{\tensor h}$).

Using equation~(\ref{eqn:deriv:point}), we see that in the limit $\gamma\to\gamma_0$,
equation~(\ref{eqn:motion:gamma}) involves the Hessian:
\begin{equation}
\label{eqn:h}
{\bf\tensor H}(\gamma_0)
= \left.\left(\frac{\partial^2{\cal U}}{\partial{\vec r}_i\partial{\vec r}_j}\right)\right|_{\gamma\to\gamma_0}
= \left.\left(\frac{\partial^2\mcalu}{\partial\mathring{\vec r}_i\partial\mathring{\vec r}_j}\right)\right|_{\gamma\to\gamma_0}
\quad,
\end{equation}
and the field of virtual forces,
\begin{equation}
\label{eqn:xi}
\xigamma(\gamma_0)
=\left.\left(-\frac{\partial^2\mcalu}{\partial\mathring{\vec r}_i\,\partial\gamma}\right)\right|_{\gamma\to\gamma_0}
\quad.
\end{equation}
In semi-condensed notation, equation~(\ref{eqn:motion:gamma}) reads:
\begin{equation}
\label{eqn:motion:gamma:0}
{\bf\tensor H}\,.
\left.\frac{{\cal D}\mathring{\bf\vec r}}{{\cal D}\gamma}\right|_{\gamma\to\gamma_0}
=\,\,\,\xigamma
\end{equation}

In order to solve equation~(\ref{eqn:motion:gamma:0}) we need to take care to eliminate
the zero modes of the Hessian.
Generically there are $d$ zero modes for a system in $d$ dimension: 
those correspond to translation invariance; their existence
indicates that a solution to equation~(\ref{eqn:motion:gamma:0}) can only be defined 
up to global translations of the particles.
There are no zero modes associated with rotations because the geometry of the simulation
cell (a parallelepiped) breaks the invariance of the problem under global rotations of the particles (at
fixed cell boundaries). 
The zero modes can thus be eliminated in the standard way by subtracting off the projection onto uniform translations from any particular solution.
Up to this invariance, the solution of~(\ref{eqn:motion:gamma:0}) reads:
\begin{equation}
\label{eqn:dr:0}
\left.\frac{{\cal D}\mathring{\bf\vec r}}{{\cal D}\gamma}\right|_{\gamma\to\gamma_0}=
{\bf\tensor H}^{-1}.\xigamma
\quad.
\end{equation}
The numerical practice which consist in, first affinely deforming the simulation cell,
then letting the system relax toward the displaced minimum, provides a direct physical
interpretation for this equation. 
$\xigamma$ is the field of forces which would result from an elementary affine deformation 
of all the particles. 
Equation~(\ref{eqn:dr:0}) shows that the non-affine displacements are just the linear
response of the system to these extra forces.

\begin{figure}
\includegraphics[width=.4\textwidth]{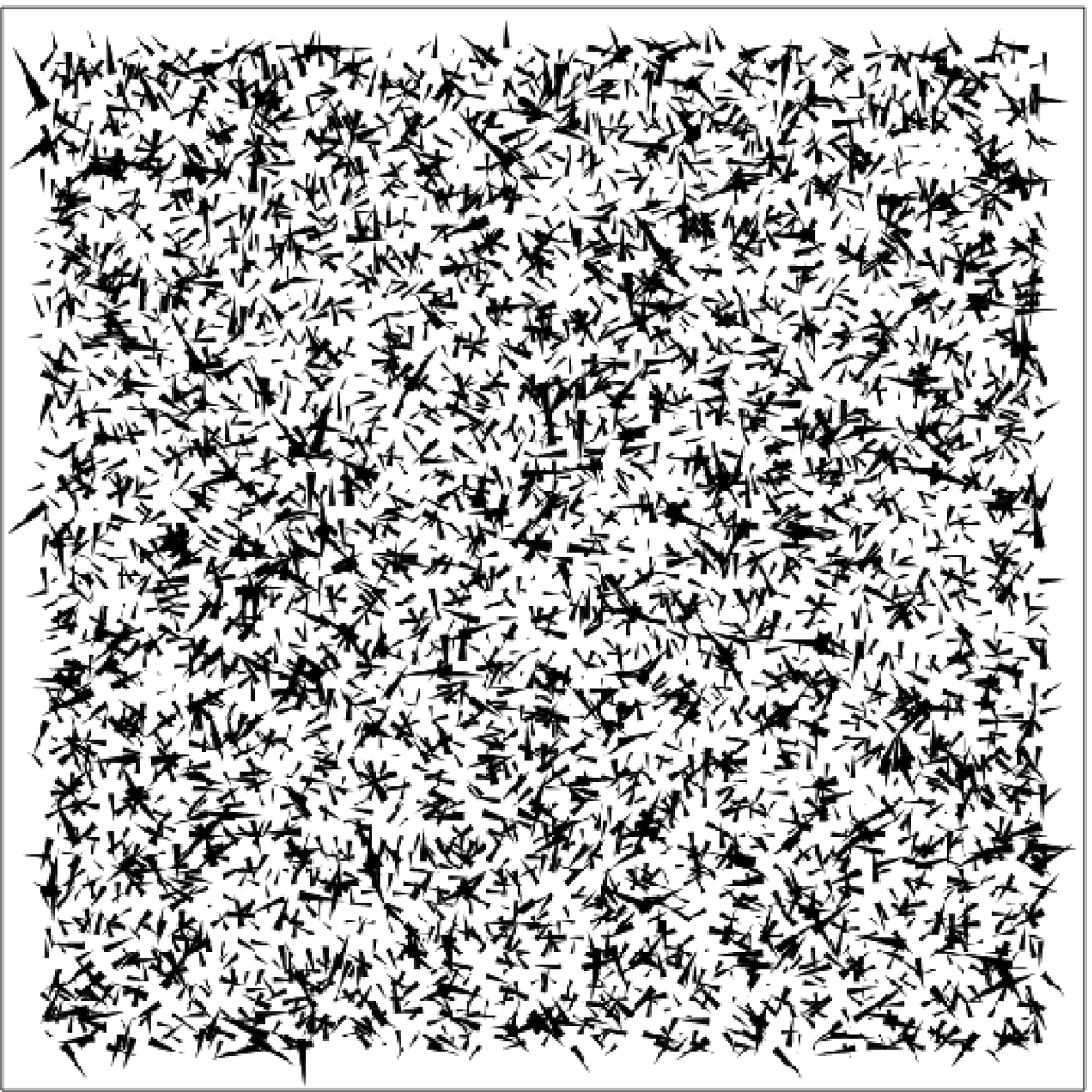}
\includegraphics[width=.4\textwidth]{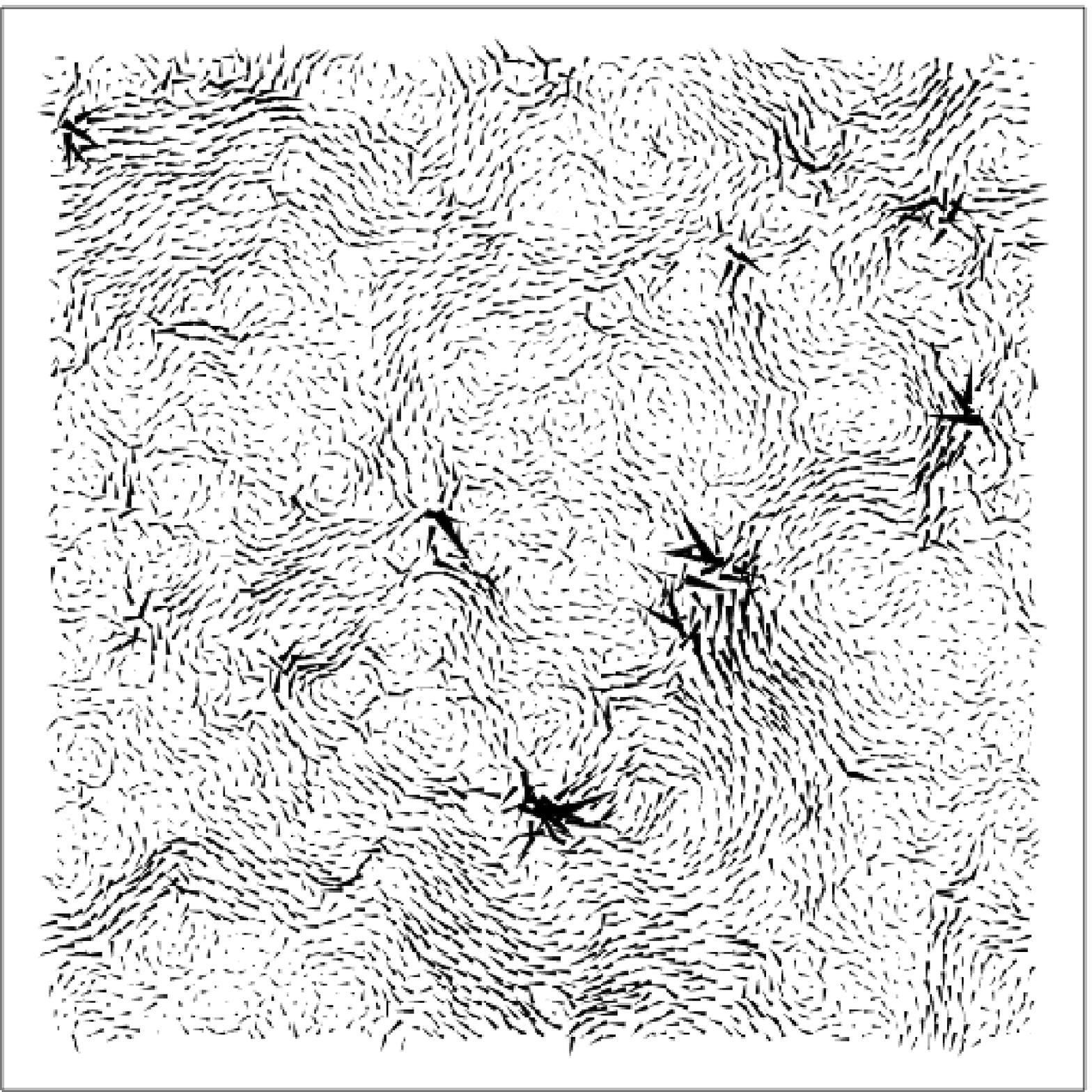}
\caption{The fields $\Xi_{i\alpha}$ (left) and 
${\cal D}\mathring{r}_{i\alpha}/{\cal D}\gamma$ (right) in pure compression.
}
\label{fig:fields:c}
\end{figure}

Examples for the fields $\xigamma$ and 
${\cal D}\mathring{\bf\vec r}/{\cal D}\gamma$ in simple
\begin{figure}
\includegraphics[width=.4\textwidth]{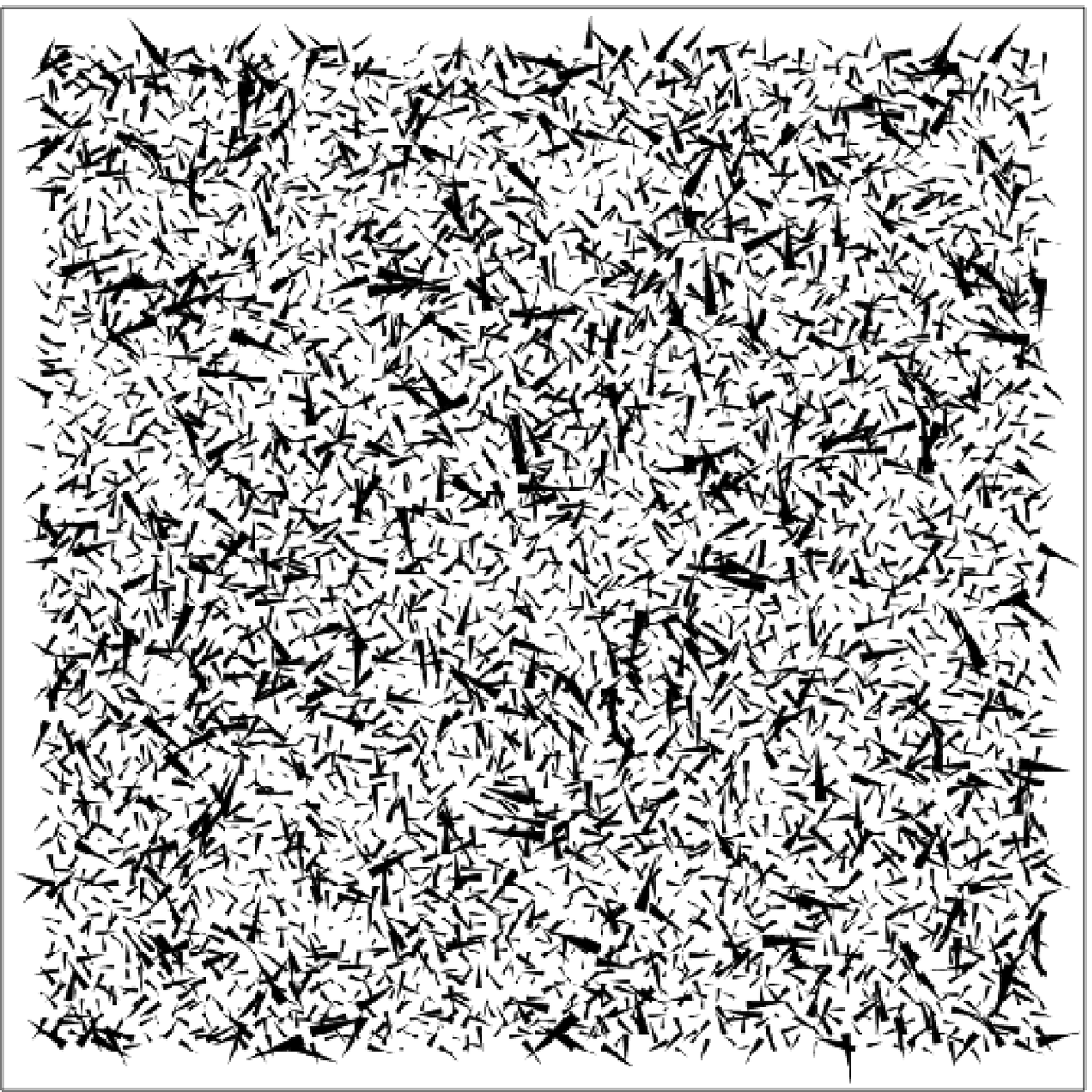}
\includegraphics[width=.4\textwidth]{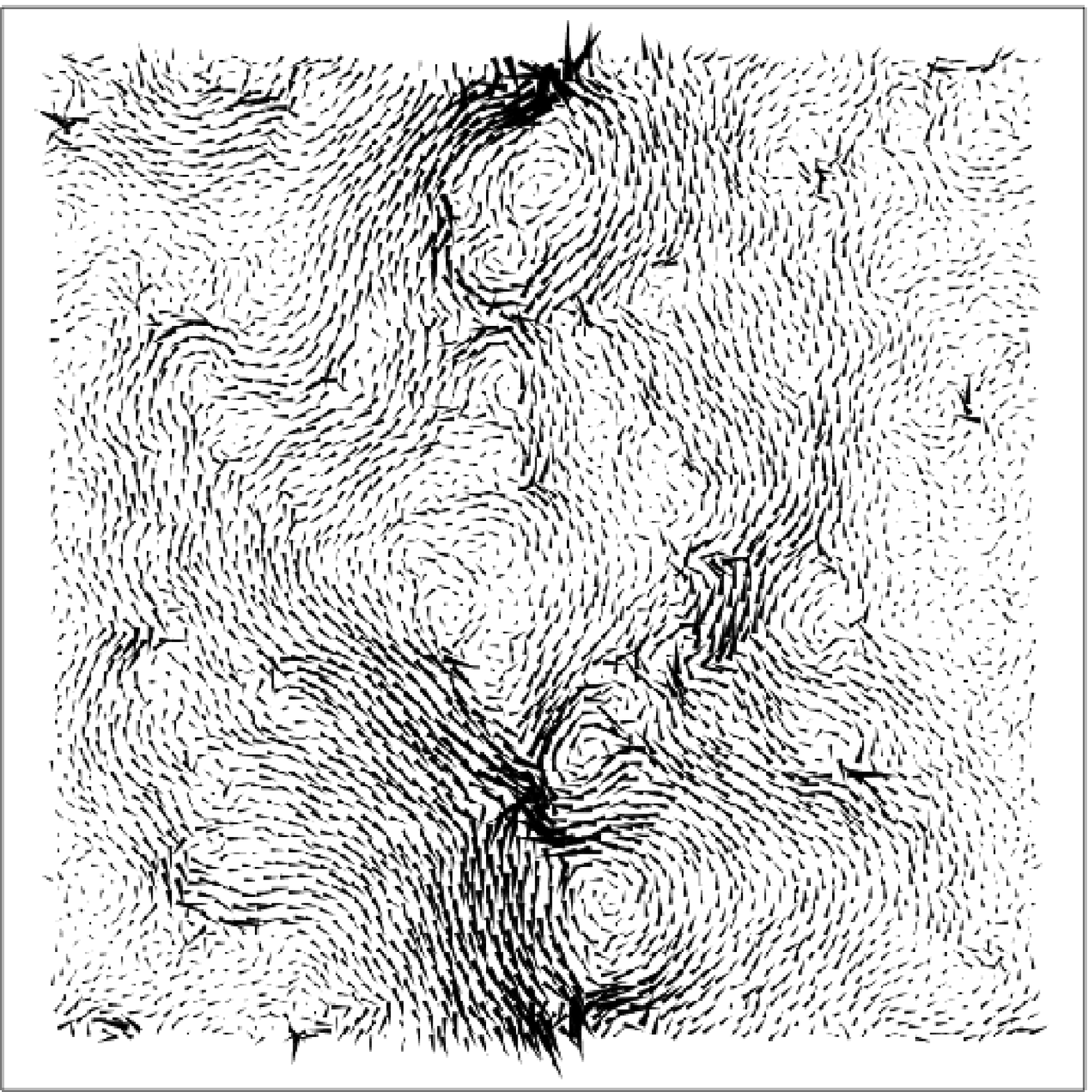}
\caption{The fields $\Xi_{i\alpha}$ (left) and 
${\cal D}\mathring{r}_{i\alpha}/{\cal D}\gamma$ (right) as defined in the text
for a typical atomistic system.
The deformation mode is simple shear.
Note the random character of $\Xi_{i\alpha}$ 
and the strong correlations in ${\cal D}\mathring{r}_{i\alpha}/{\cal D}\gamma$
}
\label{fig:fields}
\end{figure}
shear and pure compression are shown figure~\ref{fig:fields} and~\ref{fig:fields:c}.
(The details of the numerical simulation will be given in section 1.5.) 
We observe that the short-range randomness of the vector ${\bf\vec \Xi}$ 
contrasts with the large vortex-like structures displayed 
by the non-affine ``velocity'' fields ${\cal D}{\bf\mathring{\vec r}}/{\cal D}\gamma$.
We will later use the apparent disorder of the field ${\bf\vec\Xi}$ to construct
a statistical treatment of the contributions of these non-affine displacement 
to the elastic constants.
For now, let us move on to study how these non-affine fields enter microscopic
equations for the elastic constants.

Upon deformation,
the first and second derivative of the energy
${\cal U}(\gamma)=\mathring{\cal U}(\mathring{\bf\vec r}(\gamma),\gamma)$
are related to the components of the stress and the elastic constant respectively.
We will later enter a full tensorial derivation of the equations to specify this
relation. For now, we can simply differentiate ${\cal U}(\gamma)$ with respect to $\gamma$.
At first order, we obtain a projection of the stress:
\begin{equation}
\frac{{\cal D}\,{\cal U}}{{\cal D}\,\gamma}\,
=\,\frac{\partial\mathring{\cal U}}{\partial\mathring{\bf\vec r}}.\frac{{\cal D}\,\mathring{\bf\vec r}}{{\cal D}\,\gamma}
+\frac{\partial\mathring{\cal U}}{\partial\gamma}
=\frac{\partial\mathring{\cal U}}{\partial\gamma}
\end{equation}
where the second equality holds because of mechanical equilibrium.
The partial derivative which appears at the rightmost side of these equalities
corresponds to the projection of stress which would be inferred from the assumption 
that all the particles undergo affine displacements.
Since the total derivative of the energy as a function of $\gamma$
(while enforcing the condition of mechanical equilibrium)
is identical to the partial derivative, it means that 
non-affine displacements do not contribute to the \emph{first} derivative of energy with respect to $\gamma$, \emph{i.e.} the projected stress.

The situation, however, is different when the second derivative of the energy is considered.
Indeed, we have:
\begin{equation}
\frac{{\cal D}^2{\cal U}}{{\cal D}\gamma^2} 
= 
\frac{\partial^2\mcalu}{\partial\gamma^2}
+
\frac{\partial^2\mcalu}{\partial\gamma\partial \mathring{\bf\vec r}}\,.\,\frac{{\cal D}\mathring{\bf\vec r}}{{\cal D}\gamma}
= 
\frac{\partial^2\mcalu}{\partial\gamma^2}
- 
\xigamma.{\bf\tensor H}^{-1}.
\xigamma
\quad.
\end{equation}
The first term in the right hand side of this equation, $\frac{\partial^2\mcalu}{\partial\gamma^2}$,
corresponds to energy changes associated with strictly affine displacements of the particles. 
It is the Born approximation for the second derivative of the energy with respect to $\gamma$.
The last term introduces a correction to the Born approximation which results from 
the existence of non-affine displacements.
Since ${\bf\tensor H}^{-1}$ is positive definite, the corrective term is negative:
non-affine displacements reduce the second order derivative 
of the energy from its Born approximation. This is another statement of the idea
that non-affine displacement are associated with further minimization of the energy 
of the system from a tentative homogeneous deformation.
This property holds for any mode of deformation $\tensor h(\gamma)$.
It means, for example, that the non-affine corrections to the shear and compression 
moduli must be negative. 

\subsection{1.4 Tensorial forms}

In the preceding section, we considered the non-affine displacements in response to 
some prescribed mode of deformation parameterized by a scalar parameter, $\gamma$.
We now proceed to study the tensorial form of the above equations: 
derivatives of the energy functional with respect to the components of the strain 
tensor will provide analytical expressions for stress and elastic constants.

The calculations can be simplified by noting that the number of independent
components of the strain tensor is reduced by symmetry.
In particular, a useful symmetry property holds under the general assumption 
that~\cite{huang50,BH54,thurston64,BK65,Alexander98} 
the interaction potential can be written
as a function of the full set of distances between pairs of interacting particles:
${\cal U}({\bf\vec r},\tensor h)={\cal U}\left(\{r_{ij}=|\vec r_{ij}|\}\right)$,
where the index $ij$ runs overs all pairs of particles, and for each pair,
$\vec r_{ij}=\vec r_j-\vec r_i$ modulo the periodic boundary conditions. (Note that
for a periodic simulation cell,
the distance between two points depends on their positions relative 
to the boundaries of the cell: the way distances are calculated thus depends on 
the shape of the cell, hence, on $\tensor h$.)
We stress that the ability to write the energy functional in this way is not restricted to potentials
involving only pair interactions: it also holds for \emph{e.g.} 
three body interactions which depends on bond angles, as used
for silicon~\cite{SW85}, or embedded atom potentials~\cite{DB83,DB84}, as used for metals.
This formalism does not apply, however, to situations when 
rotational degrees of freedom must be taken into account as, 
for instance, in Cosserat approaches 
to granular materials~\cite{CC09,lakes95,CG95,GD04} 
or to model anisotropic objects like nematics.~\cite{GB81}
We expect a similar formal treatment to be possible in these situations,
but it will require
using a slightly more general formalism that we do not wish to address here.

Let us consider a reference configuration $\mathring{\tensor h}$ and 
a current configuration $\tensor h$. 
Suppose that $\vec r$ and $\mathring{\vec r}$ are the difference between the positions 
of two particles in both these systems of coordinates.
They are related by $\vec r=\tensor F\,.\,\mathring{\vec r}$, whence:
\begin{equation}
\vec r^2-{\mathring{\vec r}^2} = 2\ \mathring{\vec r}^T.\tensor \eta.\mathring{\vec r}
\qquad,
\label{eqn:stretch}
\end{equation}
with the Green-Saint~Venant strain tensor:~\cite{ray83,ray84,ray85}
\begin{equation}
\tensor\eta 
=\frac{1}{2}\,
\left(\tensor F^T.\tensor F-\tensor 1\right)
\quad.
\label{eqn:epsilon}
\end{equation}

We want to write the energy functional after the change of variable where
any configuration $(\{{\vec r}_i\},\tensor h)$ of the system 
is mapped onto 
$(\{\mathring{\vec r}_i\}=\tensor F^{-1}\,.\,{\vec r}_i,\tensor F)$.
Under the general assumption that the potential energy can be written as a function of
the distances $\{r_{ij}\}$ only, equation~(\ref{eqn:stretch}) can be used to write:
$$
{\cal U}({\bf\vec r},\tensor h)
= {\cal U}\left(\left\{\sqrt{\mathring{\vec r}_{ij}^2
+ 2\ \mathring{\vec r}_{ij}^T.\tensor \eta.\mathring{\vec r}_{ij}}\,\right\}\,\right)
\equiv \mcalu(\{\mathring{\vec r}_{ij}\},\tensor \eta)
\quad,
$$
Where it appears that the energy of the system is a function of $\{\mathring{\vec r}_{ij}\}$
and $\eta$ only.~\cite{huang50,BH54,thurston64,BK65,Alexander98} 
The notation $\mcalu$ is introduced here to emphasize that this functional of
$\{\mathring{\vec r}_{ij}\}$ and $\eta$
is defined after a choice of reference cell with Bravais matrix $\mathring{\tensor h}$
(whereas the functional ${\cal U}({\bf\vec r},\tensor h)$ does not depend on this choice).
A choice of $\mathring{\tensor h}$ being made, however, the energy functional depends 
on the current cell coordinates only {\it via} the symmetric
tensor $\tensor\eta$, but not {\it via} the whole tensor $\tensor h$.
We recover the separation of coordinates: 
changing $\tensor\eta$ for fixed $\{\mathring{\vec r}_i\}$
corresponds to performing an affine strain of the whole
system--the particles and the boundary; changing
$\{\mathring{\vec r}_i\}$ in the reference cell corresponds to performing
non-affine displacements of the particles.


We are now in a position to derive equations 
of motion for non-affine displacement fields in a fully tensorial form.
In any configuration, the force acting on particle $i$ is:
\begin{equation}
\label{eqn:force}
\vec f_i
= -\frac{\partial{\cal U}}{\partial\vec r_i}\Big|_{\tensor h}(\{\vec r_j\},\tensor h)
= -\frac{\partial\mcalu}{\partial\mathring{\vec r}_i}\Big|_{\tensor\eta}(\{\mathring{\vec r}_j\},\tensor\eta)\,.\,\tensor F^{-1}
\quad,
\end{equation}
and mechanical equilibrium reads:
\begin{equation}
\label{eqn:mechanical}
\forall\,i\,,\quad\quad
\frac{\partial{\cal U}}{\partial\vec r_i}\Big|_{\tensor h}(\{\vec r_j\},\tensor h)
=
\frac{\partial\mcalu}{\partial\mathring{\vec r}_i}\Big|_{\tensor\eta}(\{\mathring{\vec r}_j\},\tensor\eta)
= \vec 0
\quad.
\end{equation}
The derivatives are taken here for fixed cell coordinates.
The equation of motion for $\{\mathring{\vec r}_i\}$, is now obtained by differentiation
of~(\ref{eqn:mechanical})
with respect to the components of $\tensor\eta$:
\begin{equation}
\label{eqn:motion}
\frac{\partial^2\mcalu}{\partial\mathring{r}_i^\alpha\partial\mathring{r}_j^\beta}\,.\,
\frac{{\cal D}\mathring{r}_j^\beta}{{\cal D}\eta_{\kappa\chi}}+
\frac{\partial^2\mcalu}{\partial\mathring{r}_i^\alpha\partial\eta_{\kappa\chi}}= 0
\end{equation}
with the usual summation convention of repeated (Greek and Latin) indices.
As above, the symbol ${\cal D}$ denotes partial derivatives 
with respect to some tensorial components
while enforcing mechanical equilibrium~(\ref{eqn:mechanical}).
In the limit $\tensor\eta\to0$, equation~(\ref{eqn:motion}) involves the Hessian:
\begin{equation}
\label{eqn:h:2}
{\bf\tensor H} = 
\left.\left(\frac{\partial^2{\cal U}}{\partial{\vec r}_i\partial{\vec r}_j}\right)
\right|_{\tensor h=\mathring{\tensor h}}
=
\left.\left(\frac{\partial^2\mcalu}{\partial\mathring{\vec r}_i\partial\mathring{\vec r}_i}\right)\right|_{\tensor\eta\to\tensor0}
\quad,
\end{equation}
and the field of third order tensors,
\begin{equation}
\label{eqn:xi:2}
{\bf\vec\Xi}_{\kappa\chi} 
= \left(-\left.\frac{\partial^2\mcalu}{\partial\mathring{\vec r}_i\,\partial\eta_{\kappa\chi}}\right)\right|_{\tensor\eta\to\tensor0}
\quad.
\end{equation}
In semi-condensed notation, equation~(\ref{eqn:motion}) reads:
\begin{equation}
\label{eqn:motion:0}
{\bf\tensor H}\,.
\left.\frac{{\cal D}\mathring{\bf\vec r}}{{\cal D}\eta_{\kappa\chi} }\right|_{\tensor\eta\to\tensor0}=
{\bf\vec\Xi}_{\kappa\chi}
\end{equation}
Provided standard caution in the inversion of the Hessian, we then get,
up to translation modes:
\begin{equation}
\label{eqn:dr}
\left.\frac{{\cal D}\mathring{\bf\vec r}}{{\cal D}\eta_{\kappa\chi}}\right|_{\tensor\eta\to\tensor0}=
{\bf\tensor H}^{-1}.{\bf\vec\Xi}_{\kappa\chi}
\quad.
\end{equation}
which is the non-affine displacement field 
in response to elementary deformation along $\eta_{\kappa\chi}$.

To get further insight into the correction term in equation~(\ref{eqn:c:xi}), 
and on the physical interpretation of ${\bf\vec\Xi}$,
let us write:
\begin{equation}
\label{eqn:xi:cauchy}
{\vec\Xi}_{i,\kappa\chi}
=\left.\frac{\partial\vec f_i}{\partial\eta_{\kappa\chi}}\right|_{\tensor\eta\to\tensor0}
\quad.
\end{equation}
This equality comes after differentiation of
equation~(\ref{eqn:force}) with respect to $\tensor\eta$ and use of mechanical 
equilibrium (equation~(\ref{eqn:mechanical})).
As in our pedagogical example of one-parameter strain, the field ${\bf\vec\Xi}_{\kappa\chi}$,
can be interpreted as the forces which would result from an 
elementary affine displacement of all particles 
in the strain direction $\eta_{\kappa\chi}$. (More specifically, 
$\Delta\eta_{\kappa\chi}\,{\bf\vec\Xi}_{\kappa\chi}$, with no summation on the greek indices,
is the force resulting from a small affine transformation by $\Delta\eta_{\kappa\chi}$;
${\bf\vec\Xi}_{\kappa\chi}$ is the tangent direction to the changes of forces upon
affine transformations.)
We see that ${\vec\Xi}_{i,\kappa\chi}$ 
can be seen as the fluctuation of the force $\vec f_i$ in response to an elementary strain.
This interpretation emphasizes the random character 
of field ${\bf\vec\Xi}_{\kappa\chi}$: 
local forces are, by definition zero at mechanical equilibrium;
however, their variation under an elementary strain depends on the configuration 
of the particles with which particle $i$ interacts;
in particular,  ${\vec\Xi}_{i,\kappa\chi}$ should be zero if the conformation 
of particles surrounding $i$ is symmetric--which is the case of a Bravais crystal.
${\bf\vec\Xi}_{\kappa\chi}$ is thus a measure of the local 
asymmetry of particle configurations.

Equation~(\ref{eqn:dr}) further states that the non-affine displacement 
is nothing but the linear response
of the particles to these extra forces.

Let us now derive microscopic equations for the stress and elastic constants.
Because the total potential energy 
is a function of $\tensor\eta$ and  $\mathring{\bf\vec r}$ only
it is sufficient to characterize the elastic response of a material
using the derivatives of the potential energy with respect
to the components of $\tensor\eta$. We note, however, that 
the usual definition of the Cauchy stress and elastic stiffnesses involves
derivatives with respect to the components of the deformation gradient tensor $\tensor F$.
Derivatives with respect to the components of $\tensor\eta$ provide the so-called
thermodynamic tension and elastic constants (or also thermodynamic stiffnesses).
The definitions of thermodynamic tension, Cauchy stress, elastic constants,
elastic stiffnesses and their relation are given in appendix A. 

By definition, the thermodynamic tension is the derivative of the 
energy functional with respect to the components of $\tensor\eta$:
$$
\tensor t=\frac{1}{\mathring{V}}\,\frac{{\cal D}\mcalu}{{\cal D}\tensor\eta}
$$
Since $\tensor\eta$ is a symmetric tensor, we can conclude without further examination 
that thermodynamic tension is also symmetric.
Using mechanical equilibrium, we next have:
\begin{equation}
\label{eqn:xi:tension}
\tensor t=\frac{1}{\mathring{V}}\,\frac{{\cal D}\mcalu}{{\cal D}\tensor\eta}
=\frac{1}{\mathring{V}}\,\frac{\partial\mcalu}{\partial\tensor\eta}
\quad,
\end{equation}
As in our pedagogical example of one-parameter strain, we see here that the thermodynamic
tension is equal to the partial derivative of $\mcalu$ (here, with respect to $\tensor\eta$).
This partial derivative corresponds to changes in energy during a strictly affine
displacement of the particles: we see that the existence of non-affine displacement fields
does not appear in the expression for the thermodynamic tension.
Expression~(\ref{eqn:xi:tension}) is valid for any finite strain: this will allow us to later
differentiate this expression a second time and obtain elastic constants.
Often, though, we will use it in the limit $\tensor\eta\to0$.

Note that this expression for the thermodynamic tension provides another interpretation
of the field ${\bf\vec\Xi}$. We can write:
\begin{equation}
\label{eqn:xi:cauchy:2}
{\vec\Xi}_{i,\kappa\chi}
=-\mathring{V}\,\left.\frac{\partial t_{\kappa\chi}}{\partial\mathring{\vec r_i}}\right|_{\tensor\eta=\tensor0}
\quad,
\end{equation}
from the definitions of $\tensor t$ and ${\bf\vec\Xi}$.
This expression is to compare with equation~(\ref{eqn:xi:cauchy}).
We see that ${\vec\Xi}_{i,\kappa\chi}$ 
can be seen as a fluctuation of stress (or tension)
in response to an elementary displacement of particle $i$. This interpretation
was noted in~\cite{lutsko89}.

The elastic constants are second derivatives of the energy functional
with respect to the Green-Saint Venant strain tensor:
\begin{equation}
C_{\alpha\beta\kappa\chi}
= \frac{1}{\mathring{V}}\,
\frac{{\cal D}^2\mcalu}{{\cal D}\eta_{\alpha\beta}{\cal D}\eta_{\kappa\chi}}
\end{equation}
The derivative is here ``total'' since this is the second derivative of
the energy following deformation-induce changes of a minimum.
As we can permute the order of derivatives, elastic constants  
verify $C_{\alpha\beta\kappa\chi}=C_{\kappa\chi\alpha\beta}$, and since
$\tensor \eta$ is symmetric, 
$C_{\alpha\beta\kappa\chi}=C_{\beta\alpha\kappa\chi}=C_{\alpha\beta\chi\kappa}$.
To obtain an analytical expression for the elastic constants, it suffices to 
differentiate expression~(\ref{eqn:xi:tension}) once and take the limit $\tensor\eta\to0$ afterwards:
\begin{equation}
C_{\alpha\beta\kappa\chi}
= \frac{1}{\mathring{V}}\,
\left.\left(
\frac{\partial^2\mcalu}{\partial\eta_{\alpha\beta}\partial\eta_{\kappa\chi}}
+\frac{\partial^2\mcalu}
{\partial\mathring{\vec r_i}\partial\eta_{\alpha\beta}}\,.\,\frac{{\cal D}\mathring{\vec r_i}}{{\cal D}\eta_{\kappa\chi}}
\right)\right|_{\tensor\eta\to\tensor0}
\quad.
\end{equation}
Using equation~(\ref{eqn:dr}) and the definition of ${\bf\vec\Xi}_{\alpha\beta}$, it then comes:
\begin{equation}
\label{eqn:c:xi}
C_{\alpha\beta\kappa\chi} =
\frac{1}{\mathring{V}}\,
\left(
\left.\frac{\partial^2{\cal U}}{\partial\eta_{\alpha\beta}\partial\eta_{\kappa\chi}}\right|_{\tensor\eta\to\tensor0}
-{\bf\vec\Xi}_{\alpha\beta}.{\bf H}^{-1}.{\bf\vec\Xi}_{\kappa\chi}
\right)
\end{equation}
We recognize the first term in equation~(\ref{eqn:c:xi}) 
as the Born approximation $C^{\rm Born}_{\alpha\beta\kappa\chi}$. The contraction of the inverse
of the Hessian on components of ${\bf\vec\Xi}_{\alpha\beta}$ provides the correction
terms.

\subsection{1.5 Special case: Isotropic material in $2D$}
\label{sec:2d}


Here, we illustrate these ideas with numerical simulations of a two-dimensional
bidisperse mixture of particles interacting through a shifted
Lennard-Jones potential.~\cite{TWL+02}
Particle sizes $r_{S}=r_{L}{\sin{\frac{\pi}{10}}}/{\sin{\frac{\pi}{5}}}$
and a number ratio $N_{L}/N_{S}=\frac{1+\sqrt{5}}{4}$
are used to prevent crystallization.
The system is prepared via an initial quench
from an infinite temperature state. 
Further details regarding the numerical protocols will be found in our forthcoming dedicated numerical study~\cite{Inprep}.

The interaction potential is pair-wise additive: 
${\cal U}(\{\vec r_{ij}\}) = \sum_{ij}\,V_{ij}\,(r_{ij})$.
In this case, formal expressions for the fields ${\vec\Xi}_{i,\kappa\chi}$ 
and the Born approximation for the elastic constants can be related directly to the 
derivatives of $V_{ij}$. These expressions ((\ref{eqn:xi:micro:pair}) 
and~(\ref{eqn:born:micro:pair})) are derived in appendix B. We just recall
here equation~(\ref{eqn:born:micro:pair}):
\begin{equation}
\label{eqn:born:micro:pair:0}
C^{\rm Born}_{\alpha\beta\kappa\chi}=\frac{1}{\mathring{V}}\,\sum_{ij}\left(r_{ij}\,c_{ij}-t_{ij}\right)
\,r_{ij}\,n_{ij}^\alpha\,n_{ij}^\beta\,n_{ij}^\kappa\,n_{ij}^\chi
\end{equation}
where we have introduced the normalized vector between pairs of particles: 
$\vec n_{ij}=\frac{\vec r_{ij}}{r_{ij}}$, as well as the bond tensions and stiffnesses:
$$
t_{ij}=\frac{\partial V_{ij}}{\partial r_{ij}} \quad\text{and}\qquad 
c_{ij} = \frac{\partial^2 V_{ij}}{\partial r_{ij}^2}
\quad.
$$

The material is expected to be isotropic so that the 
elastic constants take only two independent values under permutations of indices:
$C_{\alpha\beta\kappa\chi}
=
\lambda\,\delta_{\alpha\beta}\,\delta_{\kappa\chi}
+\mu\,(\delta_{\alpha\chi}\,\delta_{\beta\kappa}+\delta_{\alpha\kappa}\,\delta_{\beta\chi})
$, which define the Lam\'e constants, $\lambda$ and $\mu$. 
Using equation~(\ref{eqn:born:micro:pair:0}) we observe that the
Born term is identical for the two Lam\'e constants $\lambda$ and $\mu$.
Introducing the notation $\vec n_{ij}= (\cos\theta_{ij},\sin\theta_{ij})$,
we find:
$$
\lambda^{\rm Born}=\mu^{\rm Born} 
= \frac{1}{\mathring{V}}\,\sum_{ij}\left(r_{ij}\,c_{ij}-t_{ij}\right)
\,r_{ij}\,\cos^2(\theta_{ij})\,\sin^2(\theta_{ij})
\quad.
$$

To calculate the Lam\'e constants, it suffices to consider 
two modes of deformation and the associated non-affine fields.
We thus consider ${\bf\vec\Xi}_s={\bf\vec\Xi}_{xy}={\bf\vec\Xi}_{yx}$ and 
${\bf\vec\Xi}_c=({\bf\vec\Xi}_{xx}+{\bf\vec\Xi}_{yy})/2$ which are
associated with pure shear and pure compression respectively.
Plots of ${\bf\vec\Xi}_{s}$ and ${\bf\vec\Xi}_{c}$, and their
corresponding displacement fields were given 
on figure~\ref{fig:fields} and~\ref{fig:fields:c} respectively.
Pure shear grants access to the sum:
$
C_{xyxy}+C_{xyyx}+C_{yxyx}+C_{yxxy}=4\,\mu
$. Pure compression grants access to the sum:
$
C_{xxxx}+C_{xxyy}+C_{yyxx}+C_{yyyy}=4\,K
$,
where $K=\lambda+\mu$ is--in two dimensions-- the compression modulus. 
With the fields ${\bf\vec\Xi}_{s}$ and ${\bf\vec\Xi}_{c}$, 
we can directly construct the non-affine correction to $\mu$:
$\tilde\mu=-{\bf\vec\Xi}_s.{\bf H}^{-1}.{\bf\vec\Xi}_s$; and to
$K$: $\tilde K=-{\bf\vec\Xi}_c.{\bf H}^{-1}.{\bf\vec\Xi}_c$.
We recall that since the Hessian is positive definite,
these corrections are necessarily negative. We thus have in all generality
that for an isotropic material, $\mu<\mu^{\rm Born}$ and $K<K^{\rm Born}$.
Using these fields, we find in this sample, for the shear and compression moduli:
$\mu^{\rm Born}\sim125$, $\tilde\mu\sim86$, whence $\mu\sim39$ and
$K^{\rm Born}\sim250$, $\tilde K\sim14$, whence $K\sim236$.
All moduli are reported in dimensionless stress units.
These values compare to the highest density systems studied by Tanguy et al~\cite{WTB+02,TWL+02}.

We note that the correction term to the compression modulus appears to be small.
This can be understood to arise from a simple property. Let us consider the case when the potential
is pairwise additive, and is homogeneous in the sense that the force
between each pair of particles is a homogeneous function of their distance: 
$F(\kappa\,r_{ij})=\kappa^\alpha\,F(r_{ij})$.
A pure compression corresponds to a global scaling of all distances:
if the forces are homogeneous, a global scaling of the distances 
preserves a state of mechanical equilibrium.
In other word, whenever the forces are homogeneous functions of the distances,
the fluctuating force field associated with pure compression, ${\bf\vec\Xi}_c$,
vanishes identically. Hence, the correction to the Born approximation vanishes
for the compression modulus: $K=\lambda+\mu=K^{\rm Born}=\lambda^{\rm Born}+\mu^{\rm Born}$. 
For a compressed Lennard-Jones system,
if the total energy is dominated by the pairs of particles which are at close distance,
the interaction is thus dominated by the repulsive power-law divergence of the 
potential. Since this power-law leads to forces which are homogeneous functions of 
the distances, this repulsive part of the Lennard-Jones potential leads to vanishing corrections
to the Born approximation. As a result the overall amplitude of the field 
${\bf\vec\Xi}_{c}$  is small, hence the small correction to $K$.
In particular, this is 
consistent with the data from Tanguy et al~\cite{WTB+02,TWL+02} 
and explains why the corrections induce 
important changes in both $\lambda$ and $\mu$ but smaller changes in their sum.

\section{2. Normal mode decomposition}
In expression~(\ref{eqn:c:xi}), the correction to elastic constants 
involves a contraction of the Hessian on the fields ${\bf\vec\Xi}_{\alpha\beta}$.
Since the Hessian is positive definite--we are perturbing around a local minimum--it 
acts as a scalar product in expressions such as:
${\bf\vec\Xi}_{\alpha\beta}.{\bf H}^{-1}.{\bf\vec\Xi}_{\kappa\chi}$.
This form of the corrections to elasticity suggests that further insight may 
be gained from a normal mode decomposition of the fields ${\bf\vec\Xi}_{\kappa\chi}$.

We denote  the eigenvectors of the Hessian by ${\bf\vec\Psi}^p$
and  the associated eigenvalues by $\lambda_p$. We introduce
a particle mass to correctly scale eigenvalues and eigenfrequencies 
and write $\lambda_p=m\,\omega_p^2$. (In our numerical simulations, the mass
is taken to be unity.)
The vector ${\bf\vec\Xi}_{\kappa\chi}$ is written as:
$$
{\bf\vec\Xi}_{\kappa\chi} = \sum_p \widehat\Xi_{p,\kappa\chi}\,{\bf\vec\Psi}^p
$$
with $\widehat\Xi_{p,\kappa\chi}={\bf\vec\Xi}_{\kappa\chi}.{\bf\vec\Psi}^p$.
Using this decomposition and equation~(\ref{eqn:dr}), the non-affine displacement fields read:
\begin{equation}
\label{eqn:dr:normal}
\left.\frac{{\cal D}\mathring{\bf\vec r}}{{\cal D}\eta_{\kappa\chi}}\right|_{\tensor\eta\to\tensor0}=
\sum_p \frac{\widehat\Xi_{p,\kappa\chi}}{\lambda_p}\,{\bf\vec\Psi}^p
\quad;
\end{equation}
the non-affine contribution to elasticity reads:
\begin{equation}
\label{eqn:munormal}
-{\bf\vec\Xi}_{\alpha\beta}.{\bf H}^{-1}.{\bf\vec\Xi}_{\kappa\chi}
= -\sum_p \frac{\widehat\Xi_{p,\alpha\beta}\,\widehat\Xi_{p,\kappa\chi}}{\lambda_p}
\quad.
\end{equation}

\begin{figure}
\includegraphics[width=.8\textwidth]{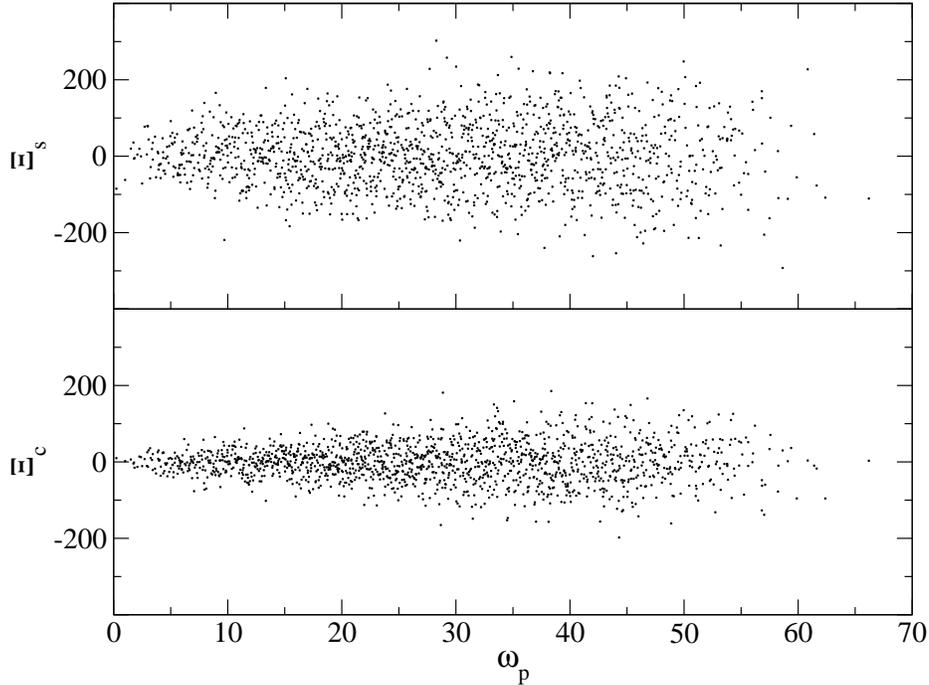}
\caption{A scatter plot of the quantities $\widehat\Xi_{p,u}$ versus 
the corresponding eigenfrequencies of the Hessian
(with $u=s$ in shear (top) and $u=c$ in compression (bottom)).
$\widehat\Xi_{p,u}$ is the projection of the field ${\bf\vec\Xi}_u$
onto the eigenmode ${\bf\vec\Psi}^{p}$ with frequency $\omega_p$.
This set of data has been obtained using one typical system of size $L=50$.
Each point makes a contribution to the non-affine corrections to the 
elastic constants as described in the text.}
\label{fig:xiScatter}
\end{figure}

As we saw from figures~\ref{fig:fields} and~\ref{fig:fields:c}, the fields ${\bf\vec\Xi}$ 
display a very random character.
The randomness of the fields $\xigamma$ can be further assessed 
by studying fluctuations of the weights of their normal mode decomposition.
To examine this question, we report, in figure~\ref{fig:xiScatter},
all values of $\widehat\Xi_{p,u}$, with $u=s,c$ corresponding 
to pure shear deformation and pure compression. 
These quantities are obtained by calculating the fields 
${\bf \vec\Xi}_{u}$ and performing a full diagonalization of the Hessian.
We see that within a frequency shell $\omega_p\in[\omega,\omega+\d\omega]$,
the values of $\widehat\Xi_{p,u}$ are scattered, so that they can indeed 
be interpreted as random variables. 
On the scatter plot, it is apparent that the distributions of $\widehat\Xi_{p,u}$ are symmetric
with respect to the horizontal axis, hence that the $\widehat\Xi_{p,u}$'s
have zero mean: this property is enforced by symmetry since each of the eigenvectors of the Hessian 
is defined up to a sign convention.
The width of the distributions of $\widehat\Xi_{p,u}$ seem to depend smoothly on $\omega$ 
and increase with $\omega$ for the most part. 
At high frequencies the scatter plots thin out because there 
are fewer and fewer eigenvectors to sample the density of states close to its upper cut-off.

On the basis of this observation, 
we make the assumption that the random fields, ${\bf\vec\Xi}_{\kappa\chi}$,
and in particular their projections on normal modes, $\widehat\Xi_{p,\kappa\chi}$,
are self-averaging quantities.
We thus introduce the correlators on frequency shells:
\begin{equation}
\label{eqn:gamma}
\Gamma_{\alpha\beta\kappa\chi}(\omega) = 
\langle\widehat\Xi_{p,\alpha\beta}\,\widehat\Xi_{p,\kappa\chi}\rangle_{\omega_p\in[\omega,\omega+\d\omega]}
\end{equation}
In this definition, the average is performed for all the projections of ${\bf\vec\Xi}_{\alpha\beta}$ 
and ${\bf\vec\Xi}_{\kappa\chi}$ on
eigenvectors with eigenfrequency $\omega_p\in[\omega,\omega+\d\omega]$.
The assumption that $\widehat\Xi_{p,\kappa\chi}$ are self-averaging 
means that the quantities $\Gamma_{\alpha\beta\kappa\chi}(\omega)$ 
are expected to converge toward well-defined functions of $\omega$ in the thermodynamic limit:
either when a large ensemble of systems of a given size is considered, 
or when one single system of a large size is considered. 

From their definition,
we observe that functions $\Gamma_{\alpha\beta\kappa\chi}(\omega)$
verify the same symmetries as the elastic constants, namely:
\begin{equation}
\label{eqn:gamma:sym}
\Gamma_{\alpha\beta\kappa\chi}=\Gamma_{\kappa\chi\alpha\beta}
=\Gamma_{\beta\alpha\kappa\chi}
\end{equation}
In the thermodynamic limit, we can finally rewrite equation~(\ref{eqn:c:xi}) as:
\begin{equation}
\label{eqn:sum}
C_{\alpha\beta\kappa\chi}=
C_{\alpha\beta\kappa\chi}^{\rm Born}
-\int_0^\infty\d\omega\,
\frac{\rho(\omega)\,{\Gamma_{\alpha\beta\kappa\chi}(\omega)}}{m\,\omega^2}\,
\quad,
\end{equation}
where $\rho$ is the density of states~\footnote{Here, $\rho$ is, somewhat unconventionally, normalized to the number of degrees of freedom per unit volume.}.
This equation is a sum rule: it relates elastic constants to the integral
of a correlator between microscopic fields.

Like the elastic constants, $\Gamma_{\alpha\beta\kappa\chi}$ 
are tensorial quantities. Therefore, if the material is 
isotropic--given the symmetry property~(\ref{eqn:gamma:sym})--
these functions can take only two independent values corresponding to the two Lam\'e
constants $\lambda$ and $\mu$:
$$
\Gamma_{\alpha\beta\kappa\chi}(\omega)=
\Gamma_\lambda(\omega)\,\delta_{\alpha\beta}\,\delta_{\kappa\chi}
+\Gamma_\mu(\omega)
\,(\delta_{\alpha\chi}\,\delta_{\beta\kappa}+\delta_{\alpha\kappa}\,\delta_{\beta\chi})
\quad.
$$
In this case, the two correlators, $\Gamma_\lambda$ and $\Gamma_\mu$ provide 
a complete description of the non-affine corrections to the Lam\'e constants
according to equation~(\ref{eqn:sum}). 

To illustrate this discussion, we return to our two-dimensional numerical study of 
two modes of deformation, simple shear and pure compression.
This protocol suffices to measure the two Lam\'e constants of our system.
The vector fields ${\bf\vec\Xi}_{s}$ and ${\bf\vec\Xi}_{c}$ 
grant direct access to the quantities $\Gamma_u=\langle\widehat\Xi_{p,u}^2\rangle$ for $u=s,c$.
Elementary algebra, similar to the calculations of section~1.5,
shows that these quantities also verify: $\Gamma_\mu=\Gamma_s$ 
and $\Gamma_K=\Gamma_\lambda+\Gamma_\mu=\Gamma_c$.

\begin{figure}
\includegraphics[width=.4\textwidth]{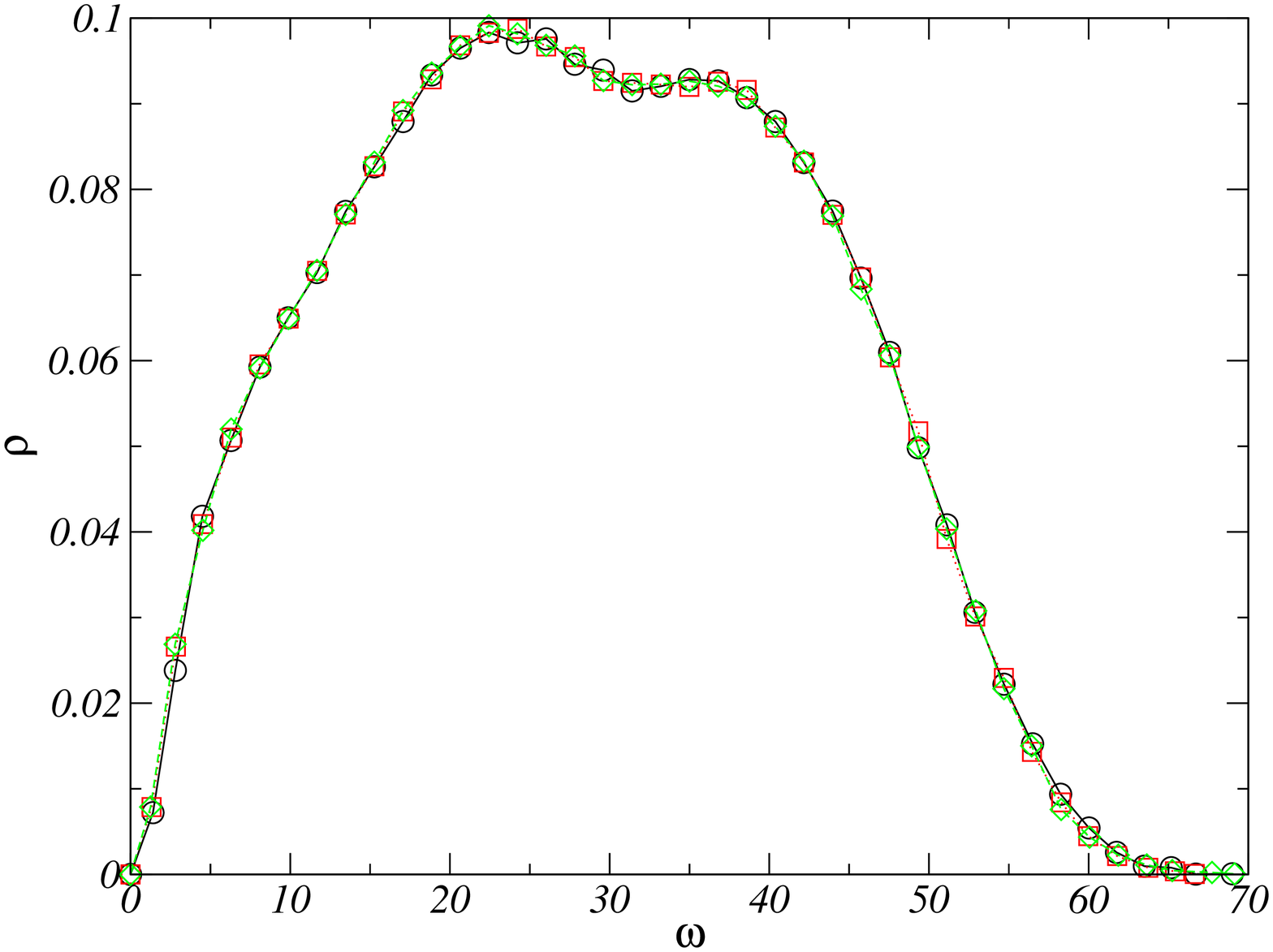}\\
\includegraphics[width=.4\textwidth]{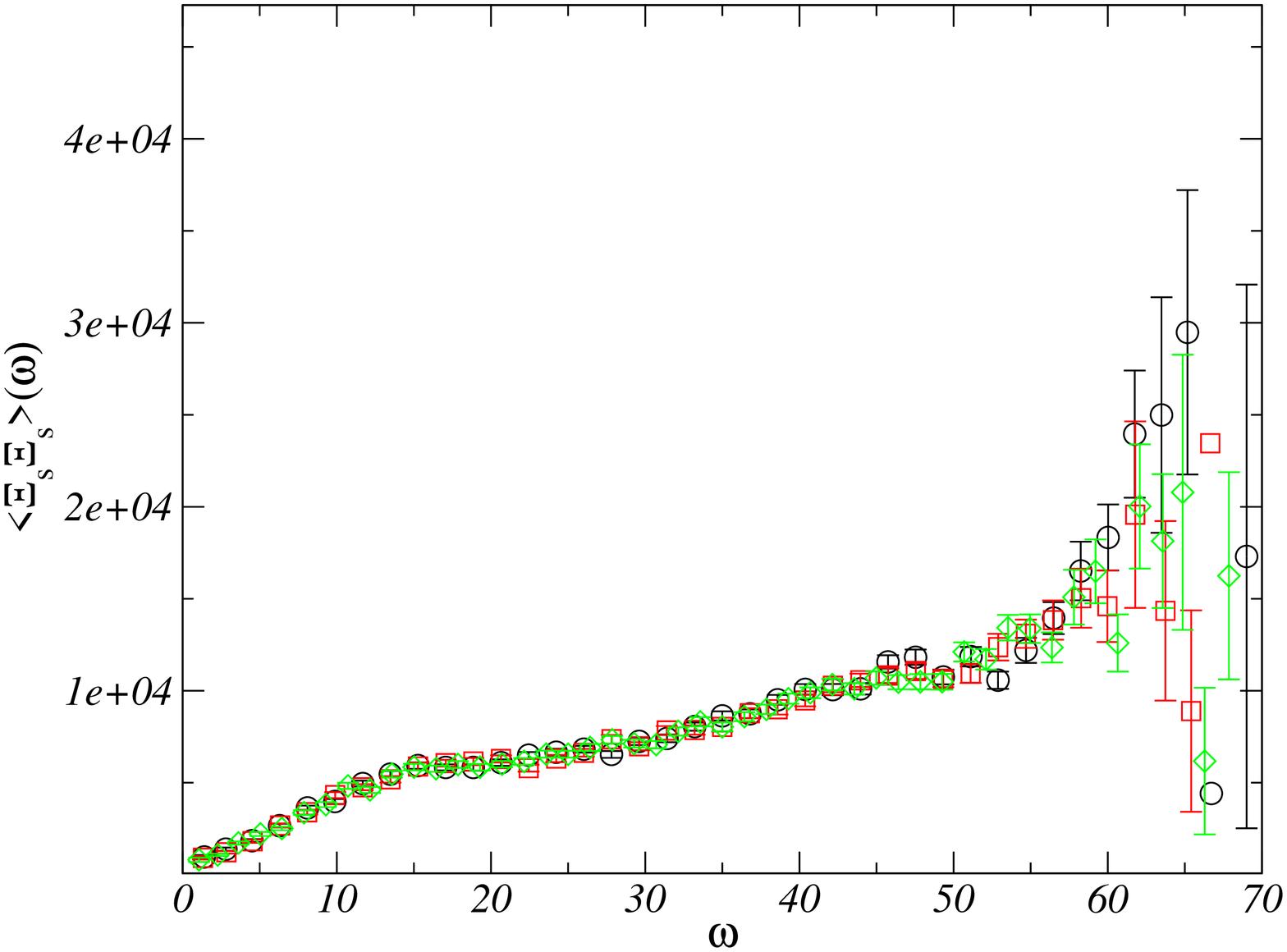}
\includegraphics[width=.4\textwidth]{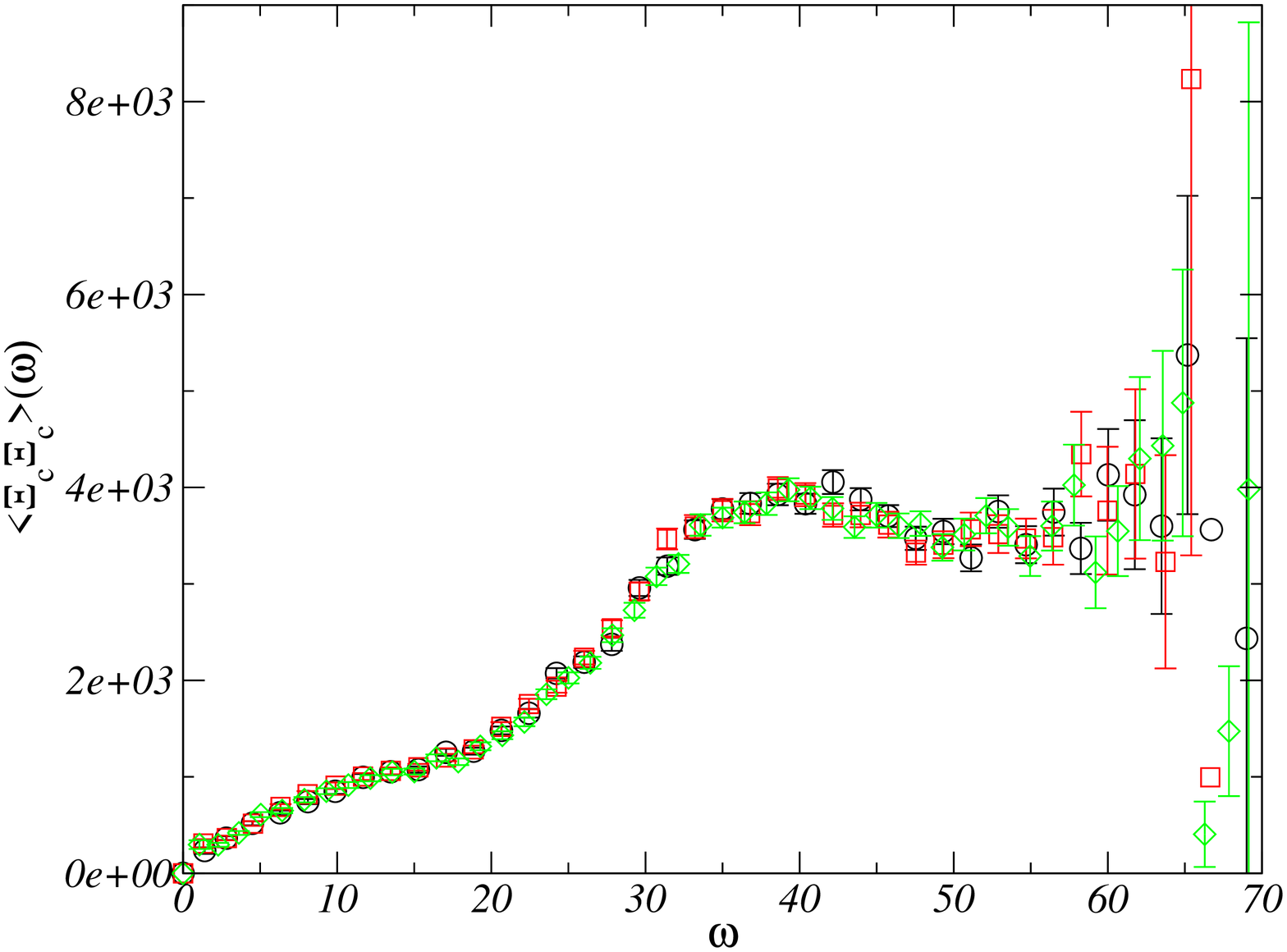}
\includegraphics[width=.4\textwidth]{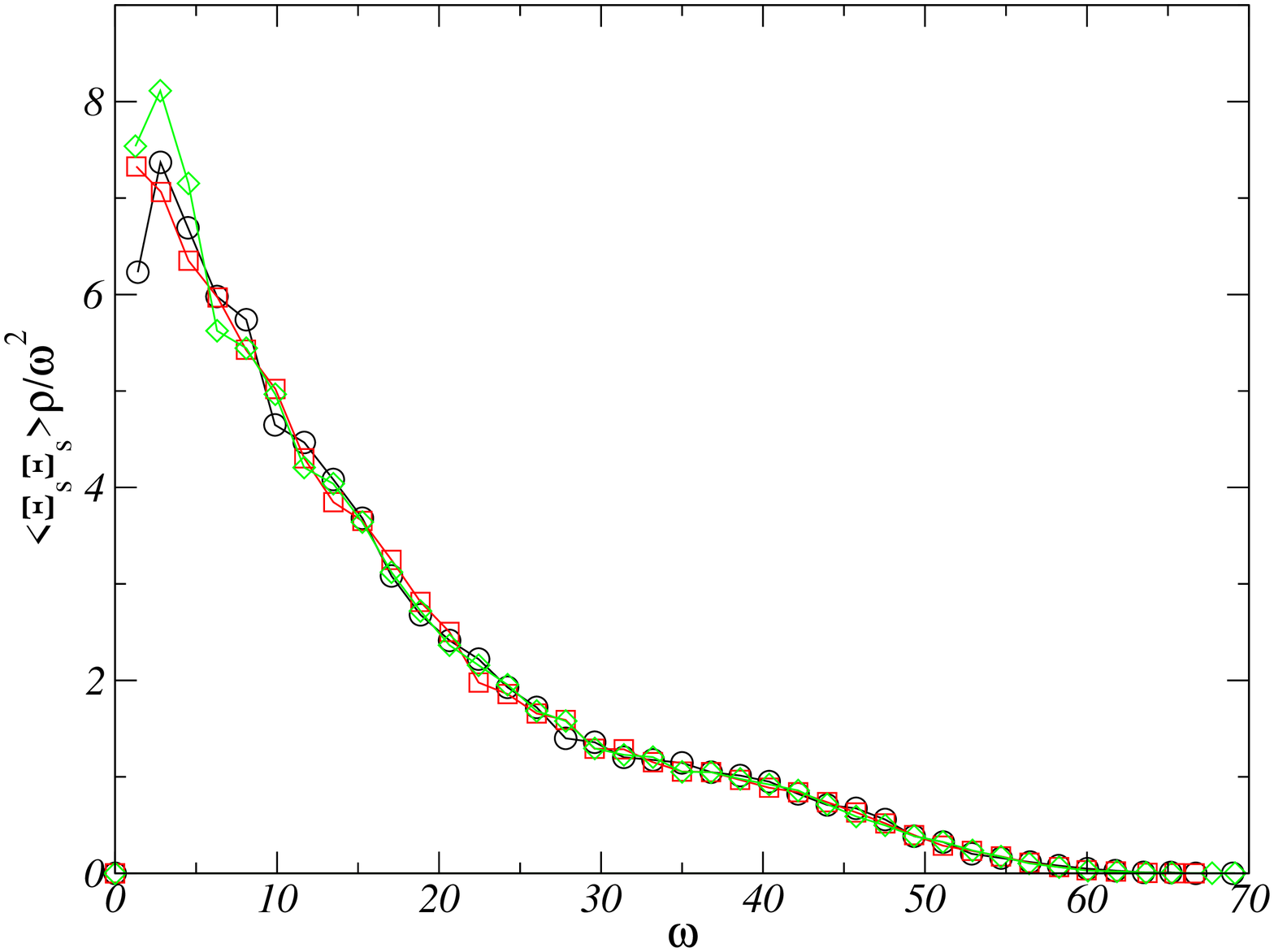}
\includegraphics[width=.4\textwidth]{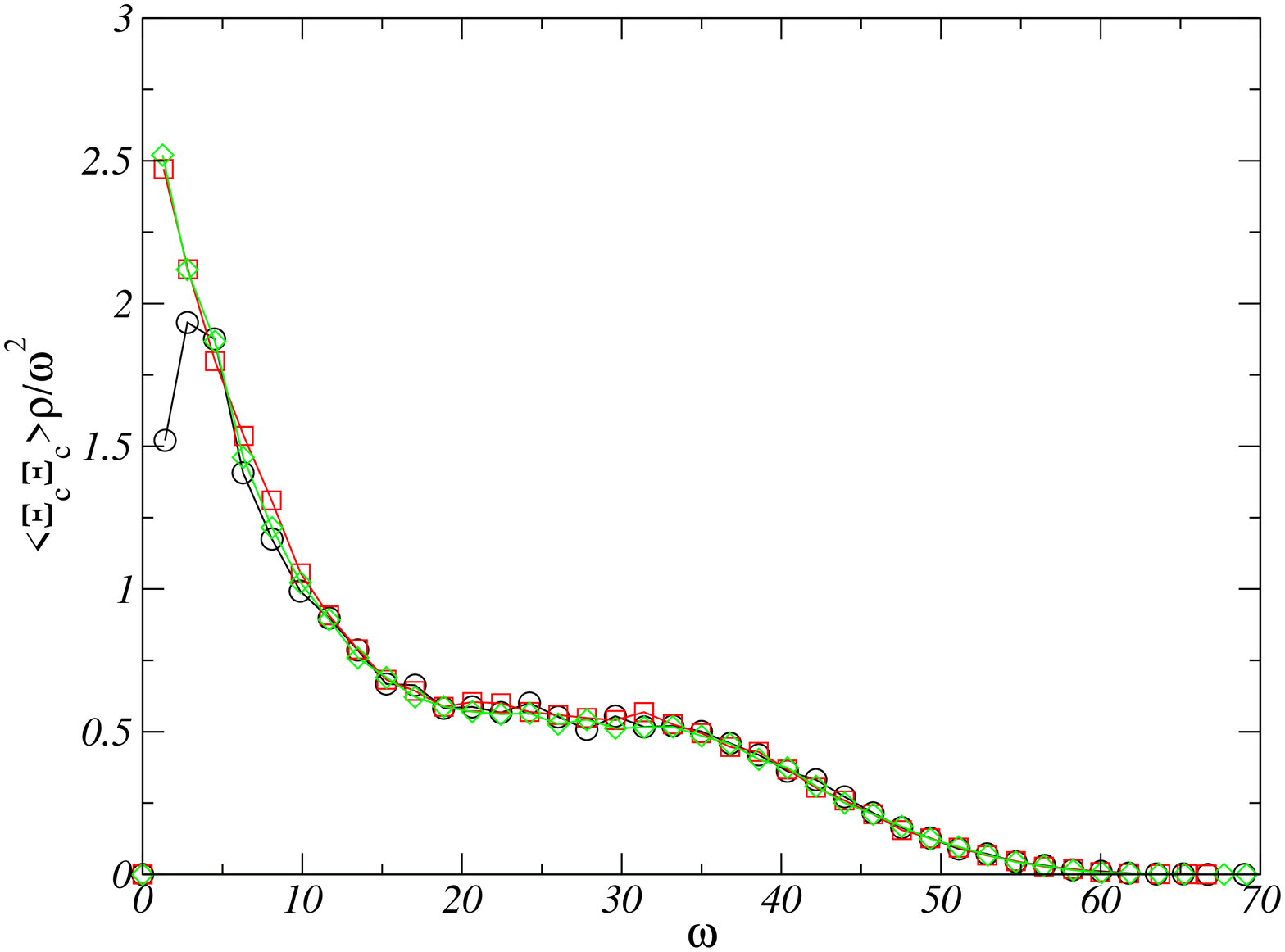}
\caption{
Ensemble averaged contribution to non-affine elasticity 
from individual bands, $[\omega,\omega+d\omega]$,
as described in the text.
Results were obtained as ensemble averages over 20 
systems of sizes $L^2$ with $L$: 20 -- circles (black), 25 -- squares (red), 
and 30 -- diamonds (green).
Convergence of the curves demonstrates that finite size effects are small.
(a) Density of states.
(b,c) Magnitude of ``noise'' field, $\langle\Xi\Xi\rangle(\omega)$, for simple shear (b) and compression (c).
(d,e) Net contribution to elasticity, $\frac{\langle\Xi\Xi\rangle\rho}{\omega^{2}}$, of each frequency band for shear (d) and compression (e).
}
\label{fig:dos}
\end{figure}

We report, in figure~\ref{fig:dos},
the density of states $\rho(\omega)$, the correlators $\Gamma_s$, $\Gamma_c$,
and the quantities
$C_u(\omega)=\rho(\omega)\,\Gamma_u(\omega)/(m\omega^2)$ (with $u = s,c$)
which measure the total contribution of the shell $[\omega,\omega+\d\omega]$ to 
the non-affine corrections. 
Three system sizes have been used in this plot:
$L=20,25,$ and $30$, where $L$ is the length of the square simulation cell in dimensionless units.
For each size, an ensemble of 20 system was used to provide statistical accuracy.

The functions $\Gamma_s=\Gamma_\mu$ and $\Gamma_c=\Gamma_K$ 
measure the variance of the distributions
of ${\widehat\Xi}_{p,s}$ and ${\widehat\Xi}_{p,c}$ respectively,
which appear on the scatter plot of figure~\ref{fig:xiScatter}.
We see that that these functions are roughly increasing as was suggested from the
observation of the scatter plot.
Important fluctuations of $\Gamma_\mu$ and $\Gamma_K$ at high frequencies 
result from the lack of statistical representation in the region where 
the density of states vanishes and where the scatter plots thin out.
For their main part, however, the curves lie on top of one another for the three system
sizes we have used: this is evidence that the continuous limit has already been reached. 
This observation is consistent with the numerics of Tanguy et al~\cite{WTB+02,TWL+02}
since the values of the elastic constants they measured did not vary much 
for the different sizes they studied, from $L\sim20$ to $L=150$.
We believe that the good convergence of these measurements results from 
the decorrelation of the fluctuating forces ${\vec\Xi}_i$ 
at very short distances: because the correlation length is short, 
the finite sizes of the systems considered here 
provide enough sampling of the distribution of ${\vec\Xi}_i$.

The density of states, and the 
$\Gamma$'s exhibit rather complicated functional
forms. 
In contrast, it is striking to us that for simple shear and pure compression $C_u(\omega)$ 
seems rather featureless; particularly in comparison with both
the density of states and the functions $\Gamma_u(\omega)$.
We have observed this for different numerical models and
different densities.~\cite{Inprep}
The monotonous decay of $C_\mu(\omega)$ and $C_K(\omega)$ seems to indicate that
a simple physical mechanism--like a transfer of elasticity from low 
to high frequency phonons--governs the non-affine corrections.
This reminds us of an analysis by Radjai and Roux of quasi-static
deformation of granular materials~\cite{RadjaiR02} 
in which the authors suggested that the non-affine velocity field 
could exhibit statistical properties similar to turbulence.
This sound like an appealing suggestion, although, at present, we have 
no evidence to further test the existence of turbulent-like energy transfers 
from large to small scales.

\section{3. Visco-elasticity}
In the above formalism, we assumed that deformation was very slowly forced
upon a piece of material, so that the relaxation of the system
toward an inherent structure was always faster than the changes
induced by the external drive. This guaranteed that the system followed
an energy minimum in configuration space.
Our analysis, however, is essentially based on a second order
expansion of the energy functional: it assumes small displacements, 
but may, in principle, apply to situations when 
the timescale of the external drive becomes relevant.

We consider here oscillatory perturbations 
of small amplitude, and assume that the system remains in the linear response 
regime around some energy minimum, far away from any catastrophic event.
We expect this situation to be of relevance to questions such as {\it e.g.\/} 
acoustic damping in granular materials~\cite{liu92,JCV99,MGJ+04,Som05} 
or viscoelasticity in dense emulsions and foams~\cite{MBW95,LGL+96,LRM+96,LL97,MLG+97}.
Near a minimum, but not exactly \emph{at} mechanical equilibrium,
the particles are subjected to non-zero forces:
an oscillatory external forcing at finite rate thus works on the system of particles.
Energy is injected into the system and is lost in dissipative mechanisms
at the microscopic scale (friction between grains or viscous dissipation between bubbles).

We will follow the usual practice of characterizing energy transfers in the
linear response regime using the components of the complex stress response, 
that is the storage and loss moduli, $G'$ and $G''$.~\cite{ferry80}
We insist that we focus here on features of $G'$ and $G''$ that
arise merely from the existence of non-affine displacement fields \emph{around a single 
minimum}. 
In an experiment--think, for example, of a foam--it is unclear whether 
sufficiently small strain amplitudes can be achieved so as to stay within the basin of attraction of a single equilibrium configuration. 
Thus measured values of $G'$ and $G''$
would also receive contributions from plasticity in addition to the microscopic visco-elasticity studied here: transitions between distinct 
inherent structures would become a relevant mechanism of energy dissipation~\cite{SLH+97,Sol98}.
It's only after a careful study of the contribution of linear response that we will be in 
a position to characterize the relative importance of these various dissipative mechanisms.
The present work is a preliminary step in this direction.

\subsection{3.1 Damped equations of motion}

In order to work at finite strain rates, we need to use Newton's equations coupled
with the deformation of our simulation cell. 
In order to prevent the system from heating up, 
a microscopic mechanism must be specified for energy dissipation.
Dissipative mechanisms vary from system to system. 
In granular packings, dissipation involves the deformation of asperities at contact
between grains. 
In foams, dissipation involves fluid flow inside a liquid phase: it is 
modeled by Durian~\cite{Dur95,Dur97} by viscous terms involving velocity differences 
between neighboring bubbles. 
In numerical simulations of structural glasses and supercooled liquids, the coupling
with an external bath is modeled either by a Nos\'e-Hoover~\cite{nose84a,nose84b,hoover85} thermostat,
by use of a Lagrange multiplier~\cite{hoover83,EM84},
or by a Berendsen thermostat ~\cite{BPV+84,AT87}.
It thus appears, after examination of these different systems, that 
energy dissipation often arises as some sort of viscous damping at the microscopic level.
We will thus limit our present discussion to the situation where dissipation enters via a 
viscous drag force applied on individual particles.
Below, we give some rule-of-thumb
estimates for what the value of 
this viscosity might be in the case of dense emulsions.


In the SLLOD algorithm, the viscous term 
is generally taken to be proportional to the peculiar velocity,~\cite{EM84} that is the velocity $\vec{\dot r}_i$
of the particle minus its velocity corresponding to the affine flow 
$\vec u(\vec r_i)=\dot{\tensor F}.\mathring{\vec r}_i=
\dot{\tensor F}.\tensor F^{-1}.{\vec r}_i$. Similarly,
in foams or granular materials, the viscous damping does not depend
on the velocity in real space but on the difference between the particles' velocity
and their surrounding environment.
We thus first write the equations of motion as:
\begin{equation}
\label{eqn:motion:damped}
m\,{\vec{\ddot r}_i}=\vec f_i-\nu\,\left(\vec{\dot r}_i-\vec u(\vec r_i)\right) 
\end{equation}
In term of $\{\mathring{\vec r}_i\}$, it follows that:
\begin{equation}
\label{eqn:motion:r00}
m\,\tensor F.\frac{\d^2{\mathring{\vec r}_i}}{\d t^2} = \vec f_i
- m\,\tensor{\ddot F}.{\mathring{\vec r}_i}
- 2\,m\,\tensor{\dot F}.\frac{\d{\mathring{\vec r}_i}}{\d t}
-\nu\,\tensor{ F}.\frac{\d{\mathring{\vec r}_i}}{\d t}
\end{equation}
We see here that a term $m\,\tensor{\ddot F}.{\mathring{\vec r}_i}$ arises
which depends on the position of the particle in the simulation cell.
Another term of similar flavor would be $\nu\,\dot{\tensor F}.\mathring{\vec r}_i$:
it does not appear in equation~(\ref{eqn:motion:r00}) because it has already been eliminated
by the above assumption that the viscous dissipation applies to the \emph{peculiar} velocities.
Such terms introduce a spatially dependent coupling between particular motion
and the cell coordinates: they cannot be allowed to enter the equations of motion
if we want to enforce translation invariance.
This was originally noted by Andersen~\cite{andersen80} and
later by Ray and Rahman~\cite{ray84}.
Within the formalism of Souza and Martins~\cite{SM97},
an equation of motion with no spatially dependent terms 
can be derived systematically from a Lagrangian.
It relies on the idea that the distance traveled by a particle should 
be calculated while removing a component of rigid rotation. 
We do not wish to detail this formalism here, but simply recall that
adding appropriate terms in the Lagrangian of the system (particles
plus simulation cell) permits eliminating space-dependent terms 
from equation~(\ref{eqn:motion:r00}).
Following these authors, we eliminate the term $m\,\tensor{\ddot F}.{\mathring{\vec r}_i}$
and focus on translation invariant equations of motion for an athermal system in SLLOD dynamics.
These equations read:
\begin{equation}
\label{eqn:motion:r0}
m\,\tensor F.\frac{\d^2{\mathring{\vec r}_i}}{\d t^2} = \vec f_i
- 2\,m\,\tensor{\dot F}.\frac{\d{\mathring{\vec r}_i}}{\d t}
-\nu\,\tensor{ F}.\frac{\d{\mathring{\vec r}_i}}{\d t}
\end{equation}


We want to account for the dynamics of the system when it is submitted
to small amounts of strain $\|\tensor F-\tensor 1\|\sim\epsilon<<1$.
To do so, we write a perturbative expansion of equation~(\ref{eqn:motion:r0})
around a known equilibrium state $\{\mathring{\vec r}_i\}$.
This expansion is written in terms of the displacements
$\{{\vec x}_i(t)=\mathring{\vec r}_i(t)-\mathring{\vec r}_i\}$,
which are also of order $\epsilon$.
At first order in $\epsilon$, the perturbed equations of motion read:
\begin{equation}
\label{eqn:motion:r0:perturbed}
m\,\frac{\d^2{{\vec x}_i}}{\d t^2} = 
\frac{\partial\vec f_i}{\partial\mathring{\vec r}_j}.{\vec x}_j
+
\frac{\partial\vec f_i}{\partial\tensor\eta}:\tensor\eta
-\nu\,\frac{\d{{\vec x}_i}}{\d t}
\end{equation}
Using the definition of ${\bf\tensor H}$ and ${\bf\ttensor\Xi}$ 
(equations~(\ref{eqn:h:2}) and~(\ref{eqn:xi:2})) it follows that:
\begin{equation}
\label{eqn:lin}
m\,\frac{\d^2{{\vec x}_i}}{\d t^2}
+\nu\,\frac{\d{{\vec x}_i}}{\d t} +
{\tensor H}_{ij}\,{\vec x}_j =
\vec\Xi_{i,\kappa\chi}\,\eta_{\kappa\chi}
\end{equation}
This equation governs the linear response of our system around a single minimum.
Starting from this equation, various limits allow us to recover more usual
expressions. For example, taking $\nu=0$ and canceling the term on the right hand side
yields the equation governing the vibration modes of the system;
canceling $m$ and $\nu$ yields the equations which governs quasi-static
response and defines the non-affine displacement fields.

To solve equation~(\ref{eqn:lin}), we perform two transformations: a Fourier transform
of the time domain, and a normal mode decomposition of the displacement field.
The Fourier components $\widetilde{\vec x}_i$ are the responses 
to perturbations of the form $\tensor\eta(t)=\widetilde{\tensor\eta}\sin(\omega t)$.
They are further decomposed onto normal modes as:
$\widetilde{\vec x}_i(\omega)=\widehat{\widetilde{ x}}_p(\omega)\,{\vec\Psi}^p_i$ and
$\widehat{\widetilde{\vec x}}_p(\omega)=\widetilde{x}_i(\omega)\,{\Psi}_i^p$. 
With these notations, we find:
\begin{equation}
-m\,\omega^2\,\widehat{\widetilde{x}}_p
+i\,\omega\,\nu\,\widehat{\widetilde{x}}_p
+m\,\omega_p^2\,\widehat{\widetilde{x}}_p = 
\widehat{\Xi}_{p,\kappa\chi}\,\widetilde\eta_{\kappa\chi}
\end{equation}
which is solved by:
\begin{equation}
\label{eqn:x:sol}
\widehat{\widetilde{x}}_p = 
-\frac{\widehat{\Xi}_{p,\kappa\chi}\,\widetilde\eta_{\kappa\chi}}{m\,\omega^2-m\,\omega_p^2-i\,\omega\,\nu}
\end{equation}

The perturbed stress is then obtained by writing at first order in $\epsilon$:
\begin{eqnarray*}
\Delta t_{\alpha\beta}
&=& \frac{1}{\mathring V}\,\left(\frac{\partial\mcalu}{\partial\eta_{\alpha\beta}}\left(\{\mathring{\vec r}_i(t)\},\tensor\eta\right)
-\frac{\partial\mcalu}{\partial\eta_{\alpha\beta}}\left(\{\mathring{\vec r}_i\},\tensor0\right)
\right)\\
&=&\frac{1}{\mathring V}\,\left(
\frac{\partial^2\mcalu}{\partial\eta_{\alpha\beta}\partial\eta_{\kappa\chi}}\,\eta_{\kappa\chi}
+
\frac{\partial^2\mcalu}{\partial\eta_{\alpha\beta}\partial\mathring{\vec r}_i}.\vec x_i
\right)
\end{eqnarray*}
Using~(\ref{eqn:x:sol}), the complex stress response in the frequency domain thus reads:
\begin{eqnarray*}
\widetilde{\Delta t}_{\alpha\beta}(\omega)
&=&
C^{\rm Born}_{\alpha\beta\kappa\chi}\,\widetilde\eta_{\kappa\chi}(\omega)
-\frac{1}{\mathring V}\,\sum_p \widehat{\Xi}_{p,\alpha\beta}\,\widehat{\widetilde{x}}_p(\omega)
\\
&=&C^{\rm Born}_{\alpha\beta\kappa\chi}\,\widetilde\eta_{\kappa\chi}(\omega)
+\frac{1}{\mathring V}\,\sum_p \frac{\widehat{\Xi}_{p,\alpha\beta}\,\widehat{\Xi}_{p,\kappa\chi}}{m\,\omega^2-m\,\omega_p^2-i\,\omega\,\nu}\,\widetilde\eta_{\kappa\chi}(\omega)
\end{eqnarray*}
It can thus be recast in the form:
$$
\widetilde{\Delta t}_{\alpha\beta}(\omega)=
G_{\alpha\beta\kappa\chi}(\omega)\,\widetilde\eta_{\kappa\chi}(\omega)
$$
with:
\begin{equation}
\label{eqn:g}
G_{\alpha\beta\kappa\chi}(\omega)
=C_{\alpha\beta\kappa\chi}^{\rm Born}
+\frac{1}{\mathring V}\,\sum_p\frac{\widehat{\Xi}_{p,\alpha\beta}\,\widehat{\Xi}_{p,\kappa\chi}}{m\,\omega^2-m\,\omega_p^2-i\,\omega\,\nu}
\end{equation}


Taking the thermodynamic limit, we can furthermore introduce
the functions $\Gamma_{\alpha\beta\kappa\chi}(\omega)$
defined in equation~(\ref{eqn:gamma}).
Equation~(\ref{eqn:g}) can then be written as an integral in frequency domain:
\begin{equation}
\label{eqn:moduli}
G_{\alpha\beta\kappa\chi}(\omega) =
C_{\alpha\beta\kappa\chi}^{\rm Born}
+\int_0^{\infty}\d\omega_p
\frac{\rho(\omega_p)\,\Gamma_{\alpha\beta\kappa\chi}(\omega_p)}
{m\,\omega^2-m\,\omega_p^2-i\,\omega\,\nu}
\quad.
\end{equation}
It is then an easy task to extract the real and imaginary part of the stress response.
Observe that we recover the Born approximation and the true  elastic constants
in the high and low frequency limits respectively.

\subsection{3.2 Overdamped systems and the relaxation spectrum}

The response moduli are often studied in the 'overdamped' limit, when the kinetic
energy remains small compared to the elastic energy.~\cite{ferry80} 
In this case, the term $m\,\omega^2$ is negligible before $m\,\omega_p^2$ 
and equation~(\ref{eqn:moduli}) can be given a simpler form. 
To do so, we introduce the timescales:
$$
\tau_p=\frac{\nu}{m\,\omega_p^2}=\frac{\nu}{\lambda_p}
$$
and write equation~(\ref{eqn:moduli}) as:
\begin{eqnarray*}
G_{\alpha\beta\kappa\chi}(\omega) &=&
C_{\alpha\beta\kappa\chi}^{\text{Born}}
-\frac{1}{2}\,\int_{-\infty}^{\infty}
\d\ln\tau_p\,
\frac{1-i\,\omega\tau_p}{1+\omega^2\,\tau_p^2}\,H(\tau_p)\\
&=& 
C_{\alpha\beta\kappa\chi}
+\frac{1}{2}\,\int_{-\infty}^{\infty}
\d\ln\tau_p\,
\frac{i\,\omega\,\tau_p}{1+i\,\omega\,\tau_p}\,H(\tau_p)\quad,
\end{eqnarray*}
where
\begin{equation}
H_{\alpha\beta\kappa\chi}(\tau)=
\sqrt{\frac{\tau}{m\,\nu}}\ \rho\left(\sqrt{\frac{\nu}{m\,\tau}}\right)\,
\Gamma_{\alpha\beta\kappa\chi}\left(\sqrt{\frac{\nu}{m\,\tau}}\right)
\end{equation}
The complex modulus $G_{\alpha\beta\kappa\chi}$ is thus naturally written
as a sum of Maxwell elements.
By definition,~\cite{ferry80} the function $H_{\alpha\beta\kappa\chi}$
is the relaxation spectrum of the system.
It is the product of the density of states
with the correlator $\Gamma_{\alpha\beta\kappa\chi}$.
The specific forms of this relation associated with a given experimental geometry
--pure shear or pure compression--derive trivially from these tensorial 
expressions.
\begin{figure}
\includegraphics[width=.9\textwidth]{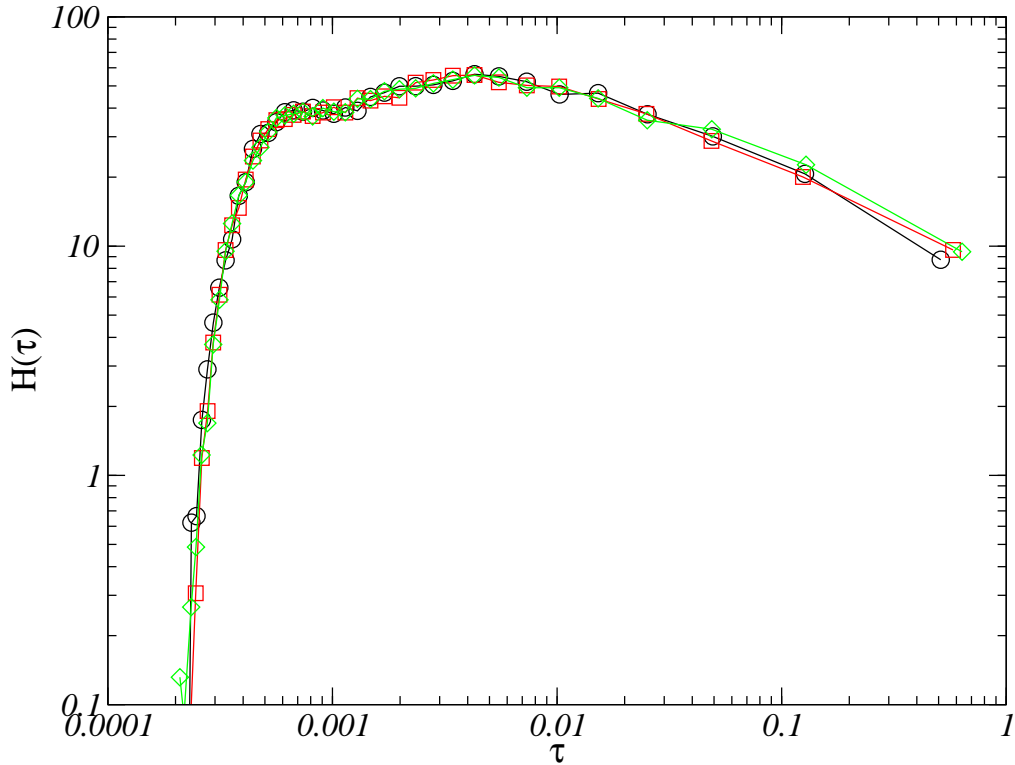}
\caption{The relaxation spectrum (as defined in the text) for simple shear.  As above, each of the three curves corresponds to a 20 fold ensemble of square systems of length, $L$: 20 -- circles (black), 25 -- squares (red), 
and 30 -- diamonds (green).}
\label{fig:relaxationSpectrum}
\end{figure}
We plot the shear relaxation spectrum for our model system in figure~\ref{fig:relaxationSpectrum}, again for each of the three ensembles of systems with lengths 20,25, and 30, demonstrating the system size independence of the results.


Since the true microscopic mechanisms of dissipation in real structural glasses remain somewhat poorly understood, we focus here on the case of a dense emulsion or foam, where the system is overdamped by construction.
To make a semi-quantitative comparison with experiments on such systems, we need to give a rough estimate for the values of the coefficient of viscosity, $\nu$, and the overall stiffness scale, $\bar{\lambda}$, in our model.
A truly quantitative connection with experiment is somewhat beyond the scope of the present work and will be explored in our dedicated numerical study~\cite{Inprep}.
We should also remind the reader that the data presented in this work for illustrative purposes is taken from a compressed system with Lennard-Jones interactions... clearly not sufficient for a fully quantitative comparison.
However, the hope is that there is a good deal of qualitative similarity in the structure of the Hessian matrix for these various classes of disordered systems, and we proceed to dimensionalize $\bar{\lambda}$ and $\nu$ in order to make a semi-quantitative comparison with experiments on packings of emulsified droplets.

A series of experiments, simulations, and theoretical works on such a system was carried out by Mason and coworkers~\cite{MBW95,LGL+96,LRM+96,MLG+97}.
The surface tension of the droplets was found to be about $\sigma\sim10~{\text{dyne}}/{\text{cm}}$, which, for well-compressed systems, should set the average scale of the interparticle stiffnesses. 
For crystalline systems, the scale of interparticle stiffness is characterized by the maximum eigenvalue of the system (corresponding to eigenvectors which are plane waves with periodicity equal to one reciprocal lattice vector), thus we scale the eigenvalues of the Hessian matrix of our system such that the maximum eigenvalues are roughly $\sigma$.
Those maximum eigenvalues of the Hessian matrix, as can be read off from the density of states, are about 5000 in our dimensionless units of stiffness, so we must have $\bar{\lambda}=\frac{[F]}{[L]}=\frac{10}{5000}=2.10^{-3}{\text{dyne}/\text{cm}}$. This sets the scales
of the second derivatives of the potential. The order of magnitude of elastic constants then
comes out by dividing $\bar\lambda$ by the average interparticule distance, that is by
the diameter of droplets in the case of an emulsion. Taking a droplet radius of a micron,
this sets the scale of elastic constants as $[C]=10~\text{dyne}/\text{cm}^2$.
Our measurement of elastic constants in the range of 100, thus corresponds to values
of order $10^3~\text{dyne}/\text{cm}^2$.

Following Durian~\cite{Dur97}, we ought to choose a value for $\nu$ such that the drag force, $F_{drag}=\nu\delta v$, between two droplets sliding past each other with relative velocity $\delta v$ generates a stress in the film which is equal to the strain rate in the film times the effective viscosity, so we must have: $\nu=F_{drag}/\delta v=\eta_{\text{eff}}\frac{R^2}{l}$ where $R$ is the typical lateral extent of the film (which, following Durian, we take to be of the order of the droplet radius itself), and $l$ is the width of the film gap.
We can now dimensionalize our units of time, $[T]=\nu/\bar{\lambda}$, using the estimate of Liu \emph{et. al.} for the effective viscosity in the film at the droplet interface of $1$cp, a droplet radius of a micron, and a film thickness of a nanometer, yielding: $[T]=5\text{s}$.
From the plot of the relaxation spectrum in figure~\ref{fig:relaxationSpectrum}, we expect the viscoelastic effects and the crossover to affine deformation to appear in the range of frequencies from around 1Hz up to about 1kHz in these systems.

\section{4. Conclusion}

In this article, we have studied  molecular displacements associated with 
deformation-induced, continuous changes of a minimum in the potential energy 
landscape of an amorphous solid.
For small amounts of deformation, the trajectory followed by the particles is smooth,
and the tangent displacements--analogous to a velocity, with strain playing the role of
an effective time--involve affine and non-affine fields. Both fields
enter at first order in the equation of motion of the particles during quasi-static 
deformation.

The non-affine fields can be calculated by inverting the Hessian on a fluctuating force
field $\bf\ttensor\Xi$, which is formally a third order tensor.
For every component $\eta_{\alpha\beta}$ of the deformation, ${\bf\vec\Xi}_{\alpha\beta}$
is the field of forces resulting on every particle from the \emph{affine} motion of its
neighborhood.
The non-affine displacement fields and the corrections they induce on elastic constants, 
can be expressed solely in terms of the Hessian and of various tensorial components
of the field ${\bf\vec\Xi}_{\alpha\beta}$. 
We next observe that the field ${\bf\vec\Xi}_{\alpha\beta}$ is weakly correlated in space,
so that it can essentially be considered as a random vector field. This randomness
is confirmed by the scattered values for the projections of ${\bf\vec\Xi}_{\alpha\beta}$
on the eigenmodes of the Hessian.

The normal-mode analysis of the field ${\bf\vec\Xi}_{\alpha\beta}$ grants 
access to the contribution of each frequency shell to non-affine corrections to elasticity.
Our numerical calculations indicate that the correlation function
$\Gamma_{\alpha\beta\kappa\chi}(\omega)=\langle\widehat{\Xi}_{p,\alpha\beta}\,\widehat{\Xi}_{p,\kappa\chi}\rangle_{\omega_p\in[\omega,\omega+\delta\omega]}$
converges toward a smooth function in the thermodynamic limit: this means that
the contribution of each frequency shell to non-affine corrections to elasticity
is self-averaging. In our view, the existence of a well-defined limit for the elastic properties
of amorphous structures, as observed by Tanguy {\it et al\/}, rest on this 
self-averaging property.

We moreover observed that the contribution of frequency shells seems to be a rather simple 
function of the frequency, as opposed to both $\Gamma_{\alpha\beta\kappa\chi}(\omega)$ and the
density of states.
This simple form of corrections to elasticity suggests that an elementary mechanism
of transfer of energy between frequency shells may be at work and determines corrections 
to elastic constants.
This is reminiscent of the observation by Radjai and Roux of turbulent like
features of non-affine displacement fields,~\cite{RadjaiR02}
although we have not determined whether turbulent-like scalings arise in our non-affine fields.
A better understanding of the mechanism underlying this transfer
through frequency shells
should in principle permit construction of approximations 
for the elastic constants of amorphous materials.

After studying elastic properties of an amorphous solid in response to quasi-static deformation,
we considered the case when the deformation rate is finite. This required us to introduce both
a molecular mass and a damping term to provide equations of motions which correspond to 
a molecular system in contact with a bath at low or zero temperature. 
We have shown that the visco-elastic
response of the solid can be written in the form of a sum of elementary damped oscillators.
The frequency spectrum--that is the distribution of timescales of elementary vibration modes-- 
is directly related to the function $\Gamma_{\alpha\beta\kappa\chi}(\omega)$.
It thus appears, that a broad spectrum of timescales arise at linear order 
 solely from the structure of any particular energy minimum in a high dimensional configuration space. 
The system remains solid since it resides in the neighborhood 
of a local minimum in configuration space at all times: this is different from the usual idea 
that visco-elastic response is associated with transitions between minima 
in the energy landscape~\cite{SLH+97,Sol98}.
In a real system, this type of dissipation may occur
simultaneously with other mechanisms: 
energy transport via phonons, dissipation resulting from anharmonic terms, or transitions
between various energy minima ({\it i.e.\/} true plasticity).
The estimation of these various contributions and their interplay
seem to us an exciting direction for future research. It will require the construction
of models, close to the experiments, where various dissipative mechanisms can
be estimated quantitatively.

Finally, we stress that the corrections to elastic constants involve $1/\omega_p^2$ 
factors as in equation~(\ref{eqn:sum}): the convergence of the integral in 
equation~(\ref{eqn:sum}) thus depends on the low frequency behavior of the functions 
$\Gamma_{\alpha\beta\kappa\chi}$ and of the density of states. 
This becomes a problem when the system develops low frequency eigenmodes 
which have a non-zero scalar product with 
the field $\bf\ttensor\Xi$: a quick inspection of equation~(\ref{eqn:munormal})
indeed shows that such localized low frequency phonons would lead to diverging terms
in the non-affine corrections to the Born approximation.
We can identify two situations when this occurs:
(i) A localized eigenvector with 
a vanishing frequency is involved whenever the local minimum occupied by the system
reaches a catastrophe.~\cite{ML97,ML99} 
This property enabled us to obtain universal scalings for the 
elastic moduli close to a shear induced catastrophe,~\cite{ML04b} and observe the divergence of the non-affine
corrections at these points. (ii) Eigenmodes 
seem to drift toward the low frequency part of the spectrum when the system approaches the 
unjamming point of Liu and Nagel.~\cite{liu98,OSL+03,WNW04,SLN05} A divergence of the non-affine corrections
to elasticity could thus arise around this point.  In this case, we expect the divergence
of the non-affine corrections to elasticity to control the unjamming transition around the $J$
point.



\acknowledgments
AL received support from the William. M. Keck Foundation, the MRSEC program of the NSF under Award No. DMR00-80034, the James. S. McDonnell Foundation, NSF Grant No. DMR-9813752, the Lucile Packard Foundation, the Mitsubishi Corporation, and the National Science Foundation under Grant No. PHY99-07949.
CM was supported under the auspices of the U.S. Department of Energy by the University of California, Lawrence Livermore National Laboratory under Contract No. W-7405-Eng-48 and would like
to acknowledge the guidance and support of V.~V. Bulatov and J.~S. Langer and the hospitality of the LLNL University Relations Program.

\appendix

\section{Appendix A: Stress and elastic constants}
\subsection{Stress and tension}

Stresses are first derivatives of an energy functional 
with respect to strains.~\cite{murnaghan51,BK65}
Since there are different definitions of the strain tensors ($\tensor\eta$ or $\tensor F$ 
or $\tensor u = \tensor F-\tensor 1$ or $\tensor e=\frac{1}{2}(\tensor u+\tensor u^T)$),
there are different ways to take this derivative.
Moreover, this derivative can be Lagrangian or Eulerian. We thus have various choices,
which leads to a number of possible definitions. Two definitions of the stress are most important:
the thermodynamic tension is the Lagrangian derivative with respect to $\tensor\eta$;
the true stress is the Eulerian derivative with respect to the deformation gradient tensor
$\tensor F$.

Let us consider some energy functional, parameterized by the cell coordinates: 
$W(\tensor h)$. This functional can represent different objects: for example, it can be
the energy $\mcalu(\{\mathring{\vec r}_i\},\tensor h)$ for fixed $\{\mathring{\vec r}_i\}$:
its strain derivatives were denoted as partial derivatives in the previous discussion;
it can also be the energy $\mcalu(\{\mathring{\vec r}_i(\tensor h)\},\tensor h)$,
provided constraints which enforce a relation $\{\mathring{\vec r}_i(\tensor h)\}$.
The formalism developed here does not depend on any specific definition of this 
energy functional but only on the existence of some function  $W(\tensor h)$.
The value of $W$ in a given reference configuration is denoted $\mathring W$.

We saw the important role played by the Green-Saint~Venant strain tensor $\tensor\eta$:
it accounts for the mapping of distances after an affine
transformation (see equation~(\ref{eqn:stretch})). 
This property permits writing the strain-dependence of the energy functional {\it via}
$\tensor\eta$ only, whenever the energy is a function of the set of distances 
$\{r_{ij}\}$ between the particles.~\cite{BK65} 
By definition, the thermodynamic 
tension~\cite{murnaghan51,thurston64,wallace72}
is conjugate to the Green-Saint~Venant strain tensor:
\begin{equation}
\label{eqn:tension}
\mathring{\tensor t} = \frac{1}{\mathring V}\frac{\partial W}{\partial \tensor\eta}
\end{equation}
Where $\mathring V$ is the volume of the reference cell.
The thermodynamic tension $\mathring{\tensor t}$ is defined after a choice of 
a reference configuration: it is a Lagrangian derivative.
It identifies, in the continuum limit, with the
second Piola-Kirchhoff stress tensor.~\cite{lubliner90,salencon01,slaughter02}
This object is known, in general, not to have a simple mechanical interpretation:
it cannot be interpreted as the tensor generating forces on surface elements.

The Cauchy stress corresponds to the usual definition of a stress:
it gives surface forces when contracted on a vector normal to a surface which is prescribed in the physical space. It can be shown that the Cauchy stress can be written as a derivative
of the energy with respect to the deformation gradient tensor:
\begin{equation}
\label{eqn:cauchy:def}
\tensor T = \frac{1}{V}\,\left.\frac{\partial W}{\partial \tensor F}\right|_{\tensor F\to\tensor 1}
\quad.
\end{equation}
Unlike the second Piola-Kirchoff stress (i.e. the thermodynamic tension)
the Cauchy stress does not depend on the choice of a reference state.
This equality holds in the limit $\tensor F\to\tensor 1$: in this sense 
it is an Eulerian derivative of the energy.
As we will show below, the second Piola-Kirchoff stress becomes identical to the Cauchy stress in the limit where the current and reference configuration are identified.

In order to relate these two definitions of the stress tensor, we need to be able to switch
between $\tensor\eta$- and $\tensor F$-derivatives.
 From the definition~(\ref{eqn:epsilon}) of the Green-Saint~Venant strain tensor, 
infinitesimal displacements read:
\begin{equation}
\d \tensor \eta = \frac{1}{2}\,
\left(\d\tensor F^T.\tensor F+\tensor F^T.\d\tensor F\right)
\label{eqn:depsilon}
\end{equation}
Since $\tensor \eta$ is symmetric, in a $d$-dimensional problem,
it has only $d(d+1)/2$ independent components, while $\tensor F$ has $d^2$.
The rotational degrees of freedom are taken into account by considering 
the antisymmetric infinitesimal displacements:~\cite{thurston64,BK65}
\begin{equation}
\d\tensor\omega = \frac{1}{2}\,
\left(\d\tensor F^T.\tensor F-\tensor F^T.\d\tensor F
\right)
\label{eqn:domega}
\end{equation}
which is the generator of infinitesimal rotations, and has $d(d-1)/2$ independent 
components. From equations~(\ref{eqn:depsilon}) and~(\ref{eqn:domega}), it comes:
\begin{equation}
\d\tensor\eta+\d\tensor\omega = \d\tensor F^T.\tensor F
\end{equation}

Using this relation, the differential form of an arbitrary function of $\tensor F$, $A(\tensor F)$, reads:
\begin{eqnarray*}
\d A &=& \Tr\left(\frac{\partial A}{\partial\tensor F}.\d\tensor F^T\right)\\
&=&\Tr\left(
\frac{\partial A}{\partial\tensor F}.\left(\d\tensor\eta+\d\tensor\omega\right).\tensor F^{-1}
\right)\\
&=&\Tr\left(
\tensor F^{-1}.\frac{\partial A}{\partial\tensor F}.\left(\d\tensor\eta+\d\tensor\omega\right)
\right)
\end{eqnarray*}
Using the property that the contraction of a symmetric with an antisymmetric tensor vanishes and the fact that $\tensor\eta$ is symmetric, we see that the derivative of a function 
$A$ with respect to $\tensor\eta$ is:
\begin{equation}
\label{eqn:deriv:eta}
\frac{\partial A}{\partial\tensor\eta}=\frac{1}{2}\,
\left[
\tensor F^{-1}.\frac{\partial A}{\partial\tensor F}
+
\frac{\partial A}{\partial\tensor F^T}.\tensor F^{-T}
\right]
\end{equation}
Incidentally, we also find from the preceding calculation that
iff $A$ only depends on $\tensor\eta$, 
hence does not vary under infinitesimal rotations, it verifies:
\begin{equation}
\label{eqn:deriv:sym}
\tensor F^{-1}.\frac{\partial A}{\partial\tensor F}
=
\frac{\partial A}{\partial\tensor F^T}.\tensor F^{-T}
\quad.
\end{equation}

With these formulae in hand, let us come back to the definition~(\ref{eqn:cauchy:def})
of the Cauchy stress.
Taking the limit $\tensor F\to\tensor 1$ (or $\tensor\eta\to0$), 
in equation~(\ref{eqn:deriv:sym}) and looking at equation~(\ref{eqn:cauchy:def}) we see that 
$\tensor T$ is symmetric. Using equation~(\ref{eqn:deriv:eta}), it appears that
it equals the thermodynamic tension \emph{in this limit\/}, 
yet in this limit only. 
We emphasize here that the symmetry of the stress tensor results
directly from the invariance of the interaction potential under 
elementary rotations: the energy functional is expected
to be a function of $\{\mathring{\vec r_i}\}$ and $\tensor\eta$ only:
its derivatives with respect to the components of $\tensor\omega$ vanish.
This is not, however, equivalent to the energy functional being invariant 
under global rotations (the Bravais cell--being a parallelepiped--is not).
We note, however, that in general $\partial W/\partial \tensor F$ (evaluated off the 
reference configuration) need not be symmetric: 
this may explain the observation of non-symmetric stresses 
in~\cite{WTB+02,TWL+02}.

Barron and Klein~\cite{BK65} provide the following relation 
between the Cauchy stress and the thermodynamic tension:
\begin{equation}
\label{eqn:cauchy}
\tensor T=\frac{1}{\det F}\,\tensor F.\mathring{\tensor t}.\tensor F^T
\quad.
\end{equation}
This relation holds for finite deformations. It can be obtained, following these authors,
after a Taylor expansion of the energy versus the different strain tensors.
We provide here a different derivation which relies on an interesting property
concerning the transport of derivatives.

As usual, a reference configuration $\mathring{\tensor h}$ is given and the system 
is strained to a current configuration $\tensor h$. In order to define the Cauchy stress around
the configuration $\tensor h$, we will need to consider $\tensor h$ as a new reference
configuration $\mathring{\tensor h}'=\tensor h$ and also consider a new current configuration $\tensor h'$.
We thus have three different sets of cell coordinates:
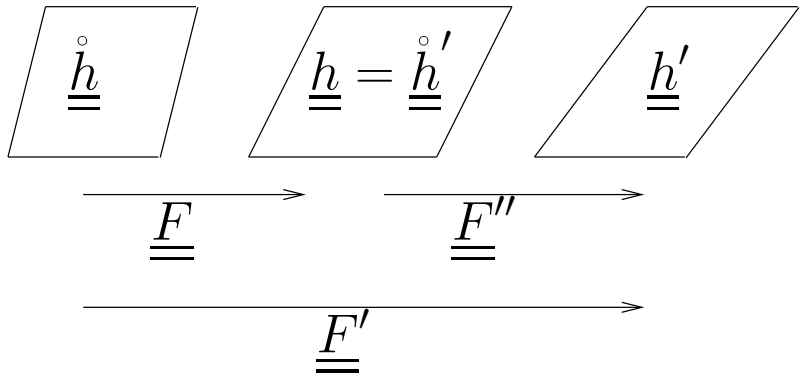
\begin{figure}[htbp]
\vspace{0.5cm}
\begin{center}
\input{transformationCartoon.pstex_t}
\end{center}
\vspace{-0.8cm}
\caption{A schematic representation of the transformations defined in the text.}
\label{fig:framesCartoon}
\end{figure}
$\mathring{\tensor h}$, $\tensor h=\mathring{\tensor h}'$ and $\tensor h'$.
We need to define strain tensors between each pair of these configuration:
the deformation gradient tensors are denoted $\tensor F''=\tensor h'.\mathring{\tensor h'}^{-1}$,
$\tensor F'=\tensor h'.\mathring{\tensor h}^{-1}$
and
${\tensor F}=\mathring{\tensor h'}.\mathring{\tensor h}^{-1}
=\tensor h.\mathring{\tensor h}^{-1}$.
These transformations are illustrated schematically in figure~\ref{fig:framesCartoon}.
Likewise, the Green-Saint~Venant tensors are denoted: 
$\tensor\eta''$, $\tensor\eta'$ and $\tensor\eta$.
We have the property: $\tensor F'=\tensor F''.{\tensor F}$.
For any arbitrary function $A(\tensor F)$, and for \emph{fixed} $\tensor F$, we can write:
\begin{eqnarray*}
\d A &=& \Tr\left(\frac{\partial A}{\partial\tensor F'}.\d\tensor{F'}^T\right)\\
&=&\Tr\left(
\frac{\partial A}{\partial\tensor F'}.{\tensor F}^{T}.\d{\tensor F''}^{T}
\right)
\end{eqnarray*}
whence,
\begin{equation}
\frac{\partial A}{\partial\tensor F''}
=
\frac{\partial A}{\partial\tensor F'}.{\tensor F}^{T}
\quad.
\label{eqn:deriv:F}
\end{equation}
Using this and equation~(\ref{eqn:deriv:eta}) it is then an easy check that:
\begin{equation}
\label{eqn:deriv:transport}
\frac{\partial A}{\partial\tensor\eta'}
=
{\tensor F}^{-1}.\frac{\partial A}{\partial\tensor\eta''}.{\tensor F}^{-T}
\end{equation}
This relation simply means that the tensorial derivative with respect
to $\tensor\eta$ transforms as a tensor under a change of reference configuration.
We just saw in equation~(\ref{eqn:deriv:F}) that this is not true of the derivative with respect to $\tensor F$.

Going back to the definition of the Cauchy stress (equation~(\ref{eqn:cauchy:def})) 
and taking the limit $\tensor h'\to\tensor h=\mathring{\tensor h}'$, we have
$$
\tensor T = \frac{1}{V}\,\left.\frac{\partial W}{\partial \tensor F''}\right|_{\tensor F'\to\tensor  F}
$$
and using 
equation~(\ref{eqn:deriv:transport}) in the limit 
$\tensor h\to\mathring{\tensor h}'$ (or $\tensor F\to\mathring{\tensor F}$), we recover
equation~(\ref{eqn:cauchy}).

It is useful to ``specialize'' the previous expressions for the situations when 
the energy, or any observable can be parameterized 
as $A(\{{\vec r}_{ij}=\tensor F.\mathring{\vec r}_{ij}\})$.
We first have:
\begin{equation}
\label{eqn:deriv:f:micro}
\frac{\partial A}{\partial \tensor F} = 
\frac{1}{2}\,\frac{\partial A}{\partial\vec r_{ij}}\,\vec r_{ij}^T.\tensor F^{-T}
\end{equation}
The factor 1/2 results from the fact that each pair is counted twice 
with the convention of implicit summation over repeated indices.
 Plugging equation~(\ref{eqn:deriv:f:micro}) (and its transpose) 
into equation~(\ref{eqn:deriv:eta}) we immediately obtain,
\begin{equation}
\label{eqn:deriv:eta:micro}
\frac{\partial A}{\partial \tensor \eta} =
\frac{1}{4}\,\tensor F^{-1}.\left(
\frac{\partial A}{\partial\vec r_{ij}}\,\vec r_{ij}^T
+\vec r_{ij}\,\frac{\partial A}{\partial\vec r_{ij}^T}
\right).\tensor F^{-T}
\end{equation}
We see from equation~(\ref{eqn:deriv:transport}) that this expression
consist of a reference-independent formula, which is transported backward
in the reference configuration as in equation~(\ref{eqn:deriv:transport}).

Specializing the functional $A$ to $A=W$, we can now provide
an expression for the Cauchy stress tensor, using either 
equation~(\ref{eqn:cauchy}) and~(\ref{eqn:deriv:eta:micro})
or alternatively ~(\ref{eqn:cauchy:def}) and~(\ref{eqn:deriv:f:micro}).
It reads:
\begin{equation}
\label{eqn:kirkwood}
\tensor T = \frac{1}{4\,V}\,\left(
\frac{\partial W}{\partial\vec r_{ij}}\,\vec r_{ij}^T
+\vec r_{ij}\,\frac{\partial W}{\partial\vec r_{ij}^T}
\right)
\quad.
\end{equation}
This is a generalization of the Kirkwood formula to the case of an arbitrary n-body interaction potential.
Note that specialization to systems with pairwise interactions immediately yields the standard expression.

Specializing the function $A$ to $A=r^\beta_{ij}$, we can furthermore obtain a useful formula.
Indeed, in the limit $\tensor F\to\tensor 1$, equation~(\ref{eqn:deriv:eta:micro}) reduces to:
\begin{equation}
\label{eqn:deriv:rij}
\left.\frac{\partial r^\beta_{ij}}{\partial\eta_{\kappa\chi}}\right|_{\tensor F\rightarrow 0} =
\frac{1}{2}\,\left(
\delta_{\beta\kappa}\, r_{ij}^\chi
+
\delta_{\beta\chi}\, r_{ij}^\kappa
\right)
\end{equation}

\subsection{Elastic stiffnesses and constants}
By definition, elastic \emph{constants} are second derivatives of the energy functional 
with respect to the Green-Saint~Venant strain tensor:
\begin{equation}
C_{\alpha\beta\kappa\chi}=\frac{1}{\mathring V}\,\left.\frac{\partial^2 W}{\partial\eta_{\alpha\beta}\partial\eta_{\kappa\chi}}\right|_{\tensor\eta\to0}
\end{equation}
In this expression, the energy functional $W$ can be any given function 
of the tensor, $\tensor \eta$. 
For example, it can be the energy functional of an atomistic system 
$W(\tensor\eta)=\mcalu(\{\mathring{\vec r}_{ij}\},\tensor \eta)$ for
fixed positions of the particles in a reference configuration. 
Derivatives of such a functional were denoted as partial derivatives of $\mcalu$ 
throughout the text. But $W$ could also be defined
as the energy of an atomistic system following deformation induced changes of a given minimum.
In this case the partial derivatives of $W$ would correspond to total derivatives for the energy 
functional $\mcalu$. The algebra presented here does not depend on these considerations
but only on the existence of a function $W(\tensor\eta)$.

Since we can commute the order of the derivatives, 
the elastic constants verify $C_{\alpha\beta\kappa\chi}=C_{\kappa\chi\alpha\beta}$, and since
$\tensor \eta$ is symmetric, 
$C_{\alpha\beta\kappa\chi}=C_{\beta\alpha\kappa\chi}=C_{\alpha\beta\chi\kappa}$.
The second order expansion of the energy with respect to the components of $\tensor\eta$ 
thus reads:
\begin{equation}
\label{eqn:second}
\frac{W-\mathring{W}}{\mathring V} = \mathring T_{\alpha\beta}\,\eta_{\alpha\beta}
+\frac{1}{2}\,C_{\alpha\beta\kappa\chi}\,\eta_{\alpha\beta}\,\eta_{\kappa\chi}
+O(\eta_{\alpha\beta}^3)
\end{equation}
where $\mathring T_{\alpha\beta}=\mathring T_{\beta\alpha}
=(1/\mathring V)\,\left.\partial W/\partial F_{\alpha\beta}\right|_{\tensor\eta\to0}$ 
is the stress in the reference configuration. Using~(\ref{eqn:tension}) and~(\ref{eqn:second}),
we find for the thermodynamic tension:
\begin{align}
\label{eqn:tension:exp}
\mathring t_{\alpha\beta}
&=\mathring T_{\alpha\beta}+C_{\alpha\beta\kappa\chi}\,\eta_{\kappa\chi}
+O(\eta_{\kappa\chi}^2)\\
&=\mathring T_{\alpha\beta}+C_{\alpha\beta\kappa\chi}\,u_{\kappa\chi}
+O(u_{\kappa\chi}^2)
\end{align}

The elastic constant should not be confounded with the quantities which enter 
the stress-strain relations--hence, the wave equation. These are
the elastic stiffnesses, $c_{\alpha\beta\kappa\chi}$, and appear in the expansion of 
the energy with respect to tensor $\tensor u=\tensor F-\tensor 1$:~\cite{BK65}
\begin{equation}
\label{eqn:second:u}
\frac{W-\mathring{W}}{\mathring V} = \mathring T_{\alpha\beta}\,u_{\alpha\beta}
+\frac{1}{2}\,c_{\alpha\beta\kappa\chi}\,u_{\alpha\beta}\,u_{\kappa\chi}
+O(u_{\alpha\beta}^3)
\end{equation}
The relation of the elastic stiffnesses to the elastic constants
is obtained by replacing the definition~(\ref{eqn:epsilon}) of $\tensor\eta$ 
in equation~(\ref{eqn:second}) and expanding in terms of $\tensor u$:~\cite{BK65}
\begin{equation}
\label{eqn:stiffness}
c_{\alpha\beta\kappa\chi}=C_{\alpha\beta\kappa\chi}
-\frac{1}{2}\,\left(
2\,\delta_{\kappa\chi}\,\mathring T_{\alpha\beta}
-\delta_{\beta\chi}\,\mathring T_{\alpha\kappa}
-\delta_{\alpha\chi}\,\mathring T_{\beta\kappa}
-\delta_{\beta\kappa}\,\mathring T_{\alpha\chi}
-\delta_{\alpha\kappa}\,\mathring T_{\beta\chi}
\right)
\quad.
\end{equation}
Note that this expression indicates that elastic stiffnesses do not enjoy the full
symmetry of the elastic constants. Instead, they verify:~\cite{BK65}
$$
c_{\alpha\beta\kappa\chi}-c_{\kappa\chi\alpha\beta}
=\mathring T_{\kappa\chi}\,\delta_{\alpha\beta}-
\mathring T_{\alpha\beta}\,\delta_{\kappa\chi}
\quad.
$$

The Cauchy stress can now be obtained from~(\ref{eqn:tension:exp}) by use of
equations~(\ref{eqn:cauchy}) and~(\ref{eqn:stiffness}):~\cite{BK65}
\begin{equation}
T_{\alpha\beta}
=\mathring T_{\alpha\beta}+
\left(c_{\alpha\beta\kappa\chi}
+\mathring T_{\alpha\chi}\,\delta_{\beta\kappa}
-\mathring T_{\alpha\beta}\,\delta_{\kappa\chi}
\right)\,u_{\kappa\chi}+O(\eta_{\kappa\chi}^2)
\end{equation}

Before proceeding to the derivation of an expression for the Born term,
let us derive a general formula for second derivatives with respect to the components
of $\tensor\eta$. We first write equation~(\ref{eqn:deriv:eta}) with indices explicitly expressed:
\begin{equation}
\frac{\partial A}{\partial\eta_{\kappa\chi}}=\frac{1}{2}\,
\left[
\left(\tensor F^{-1}\right)_{\kappa\rho}\,\frac{\partial A}{\partial F_{\rho\chi}}
+
\left(\tensor F^{-1}\right)_{\chi\rho}\,\frac{\partial A}{\partial F_{\rho\kappa}}
\right]
\end{equation}
The derivative of this expression with respect to $\eta_{\alpha\beta}$ reads:
$$
\frac{\partial^2 A}{\partial\eta_{\alpha\beta}\partial\eta_{\kappa\chi}}=\frac{1}{2}\,
\left[
\frac{\partial\left(\tensor F^{-1}\right)_{\kappa\rho}}{\partial\eta_{\alpha\beta}}\,
\frac{\partial A}{\partial F_{\rho\chi}}
+
\frac{\partial\left(\tensor F^{-1}\right)_{\chi\rho}}{\partial\eta_{\alpha\beta}}\,
\frac{\partial A}{\partial F_{\rho\kappa}}
+
\left(\tensor F^{-1}\right)_{\kappa\rho}\,\frac{\partial^2 A}{\partial\eta_{\alpha\beta}\partial F_{\rho\chi}}
+
\left(\tensor F^{-1}\right)_{\chi\rho}\,\frac{\partial^2 A}{\partial\eta_{\alpha\beta}\partial F_{\rho\kappa}}
\right]
$$
we then use the property that in the limit $\tensor\eta\to0$,
$$
\left.\frac{\partial \left(\tensor F^{-1}\right)_{\kappa\chi}}{\partial F_{\alpha\beta}}\right|_{\tensor F\rightarrow 1}
\to -\delta_{\alpha\kappa}\,\delta_{\beta\chi}
$$
by definition, whence we see that 
$$
\left.\frac{\partial \left(\tensor F^{-1}\right)_{\kappa\chi}}{\partial\eta_{\alpha\beta}}\right|_{\tensor \eta\rightarrow 0}
\to -\frac{1}{2}\,\left(\delta_{\alpha\kappa}\,\delta_{\beta\chi}
+\delta_{\beta\kappa}\,\delta_{\alpha\chi}\right)
$$
after application of equation~(\ref{eqn:deriv:eta}).
In the limit $\tensor\eta\to0$, we thus have for the second derivatives, after a further application of equation~(\ref{eqn:deriv:eta}):
\begin{align}
\frac{\partial^2 A}{\partial\eta_{\alpha\beta}\partial\eta_{\kappa\chi}}
&=-\frac{1}{4}\,
\left(
\delta_{\alpha\kappa}\,\frac{\partial A}{\partial F_{\beta\chi}}
+\delta_{\beta\kappa}\,\frac{\partial A}{\partial F_{\alpha\chi}}
+\delta_{\alpha\chi}\,\frac{\partial A}{\partial F_{\beta\kappa}}
+\delta_{\beta\chi}\,\frac{\partial A}{\partial F_{\alpha\kappa}}
\right) \nonumber \\
+&\frac{1}{4}\,\left[
\frac{\partial^2 A}{\partial F_{\alpha\beta}\partial F_{\kappa\chi}}
+
\frac{\partial^2 A}{\partial F_{\alpha\beta}\partial F_{\chi\kappa}}
+
\frac{\partial^2 A}{\partial F_{\beta\alpha}\partial F_{\kappa\chi}}
+
\frac{\partial^2 A}{\partial F_{\beta\alpha}\partial F_{\chi\kappa}}
\right]
\end{align}

We can now provide microscopic expressions for the Born approximation to
elastic constants. We take the energy functional $W$ to be the energy $\mcalu$
for some fixed positions of the particles in a reference frame. 
Using equation~(\ref{eqn:deriv:eta:micro}) and~(\ref{eqn:deriv:f:micro})
it is an easy task to write the Born term as:
\begin{align}
\label{eqn:born:micro}
C^{\rm Born}_{\alpha\beta\kappa\chi}
&= 
\frac{1}{16\,\mathring V}\,\left(
\frac{\partial^2{\cal U}}{\partial r_{ij}^\alpha\,\partial r_{kl}^\kappa}
\,r_{ij}^\beta\,r_{kl}^\chi
+
\frac{\partial^2{\cal U}}{\partial r_{ij}^\alpha\,\partial r_{kl}^\chi}
\,r_{ij}^\beta\,r_{kl}^\kappa
+
\frac{\partial^2{\cal U}}{\partial r_{ij}^\beta\,\partial r_{kl}^\kappa}
\,r_{ij}^\alpha\,r_{kl}^\chi
+
\frac{\partial^2{\cal U}}{\partial r_{ij}^\beta\,\partial r_{kl}^\chi}
\,r_{ij}^\alpha\,r_{kl}^\kappa
\right) \nonumber \\
&- \frac{1}{4}\,\left(
\delta_{\alpha\kappa}\,\mathring T_{\beta\chi}
+\delta_{\alpha\chi}\,\mathring T_{\beta\kappa}
+\delta_{\beta\kappa}\,\mathring T_{\alpha\chi}
+\delta_{\beta\chi}\,\mathring T_{\alpha\kappa}
\right)
\end{align}

In order to obtain a similar expression for 
${\vec\Xi}_{i,\kappa\chi}$, it is convenient to write:
$$
{\Xi}^\alpha_{i,\kappa\chi} = \sum_j\,{\Xi}^\alpha_{ij,\kappa\chi}
$$
with
\begin{equation}
{\Xi}^\alpha_{ij,\kappa\chi}
= -\left.\frac{\partial^2\mcalu}{\partial\mathring{r}^\alpha_{ij}
\,\partial\eta_{\kappa\chi}}\right|_{\tensor\eta\to\tensor0}
= -\left.\frac{\partial^2{\cal U}}{\partial{r}^\alpha_{ij}
\,\partial F_{\kappa\chi}}\right|_{\tensor F\to\tensor 1}
\end{equation}
The pair contributions ${\vec\Xi}_{ij,\kappa\chi}$ are then easily expressed
in terms of pair contributions to the Hessian:
$$
{\Xi}^\alpha_{ij,\kappa\chi}
=
-\frac{1}{2}\,\frac{\partial^2{\cal U}}{\partial{r}^\alpha_{ij}\partial{r}^\beta_{kl}}\,
\left.\frac{\partial{r}^\beta_{kl}}{\partial F_{\kappa\chi}}\right|_{\tensor F\to\tensor 1}
$$
Then, either using equation~(\ref{eqn:deriv:rij}) or
applying equation~(\ref{eqn:deriv:eta:micro}) on $\vec f_i$, 
we find for ${\vec\Xi}_{ij,\kappa\chi}$ an expression which is analogous to Kirkwood's
formula for the stress:
\begin{equation}
\label{eqn:xi:micro}
{\Xi}^\alpha_{ij,\kappa\chi} = -\frac{1}{4}\left(
\frac{\partial^2{\cal U}}{\partial r^\alpha_{ij}\partial r_{kl}^\kappa}\,r_{kl}^\chi
+
r_{kl}^\kappa\,\frac{\partial^2{\cal U}}{\partial r^\alpha_{ij}\partial r_{kl}^\chi}
\right)
\quad.
\end{equation}
This general formula permits relating ${\vec\Xi}_{i,\kappa\chi}$ to 
elementary contributions
 $\frac{\partial^2{\cal U}}{\partial\vec r_{ij}\partial \vec r_{kl}}$ of pairs to
the Hessian. 

We furthermore notice the similarity between equation~(\ref{eqn:xi:micro})
and equation~(\ref{eqn:born:micro}). It allows writing the following 
expression for the Born term:
\begin{align}
\label{eqn:born:micro:xi}
C^{\rm Born}_{\alpha\beta\kappa\chi}
&= 
\frac{1}{4\,\mathring V}\,\left(
{\Xi}^\alpha_{ij,\kappa\chi}\,r_{ij}^\beta
+
{\Xi}^\beta_{ij,\kappa\chi}\,r_{ij}^\alpha
\right)
&- \frac{1}{4}\,\left(
\delta_{\alpha\kappa}\,\mathring T_{\beta\chi}
+\delta_{\alpha\chi}\,\mathring T_{\beta\kappa}
+\delta_{\beta\kappa}\,\mathring T_{\alpha\chi}
+\delta_{\beta\chi}\,\mathring T_{\alpha\kappa}
\right)
\end{align}

\section{Appendix B: Microscopic expression for pair interaction potentials}
\label{app:pair}
We provide here specific expressions for the case when the interaction
potential can be written as:
$$
{\cal U}(\{\vec r_{ij}\}) = \sum_{ij}\,V_{ij}\,(r_{ij})
\quad.
$$
In this case, the force on individual bonds reads:
\begin{equation}
\label{eqn:force:ij}
\vec f_{ij}=-\frac{\partial V_{ij}}{\partial\vec r_{ij}}=
-\frac{\partial V_{ij}}{\partial r_{ij}}\,\vec n_{ij}
\end{equation}
with, 
$$
\vec n_{ij}=\frac{\vec r_{ij}}{r_{ij}}
$$
Introducing the bond tensions and stiffnesses,
$$
t_{ij}=\frac{\partial V_{ij}}{\partial r_{ij}} \quad\text{and}\qquad 
c_{ij} = \frac{\partial^2 V_{ij}}{\partial r_{ij}^2}
\quad,
$$
the components of the Hessian 
$$
{\tensor H}_{ij} =
\frac{\partial^2{\cal U}}{\partial{\vec r}_i\partial\vec r}_j
$$
can be expressed in terms of:
$$
{\rm\tensor M}_{ij} = \frac{\partial^2 V_{ij}}{\partial \vec r_{ij}\partial \vec r_{ij}} = \left(c_{ij}-\frac{\,t_{ij}}{r_{ij}}\right)\,\vec n_{ij}\,\vec n_{ij}+\frac{\,t_{ij}}{r_{ij}}\,\tensor 1
\quad,
$$
The elements of the Hessian are then,
${\rm\tensor H}_{ij} = -{\rm\tensor M}_{ij}
$
for the off-diagonal terms
and ${\rm\tensor H}_{ii}=\sum_{j\ne i} {\rm\tensor M}_{ij}$ for the diagonal terms.

To obtain an expression for the field ${\bf\vec\Xi}_{\kappa\chi}$,
we write:
\begin{equation}
\label{eqn:xii:xiij}
{\Xi}^\alpha_{i,\kappa\chi} = \sum_j {\Xi}^\alpha_{ij,\kappa\chi}
\end{equation}
with (no implicit sum on $i$ and $j$):
$$
{\Xi}^\alpha_{ij,\kappa\chi}
= -{\rm M}_{ij,\alpha\beta}\,\frac{\partial r_{ij}^\beta}{\partial\eta_{\kappa\chi}}
= -\frac{1}{2}\,{\rm M}_{ij,\alpha\beta}\,\left(\delta_{\beta\kappa}\,r_{ij}^\chi+\delta_{\beta\chi}\,r_{ij}^\kappa\right)
\quad,
$$
whence,
\begin{equation}
{\Xi}^\alpha_{ij,\kappa\chi} =
-\left(r_{ij}\,c_{ij}-t_{ij}\right) n_{ij}^\alpha\,n_{ij}^\kappa\,n_{ij}^\chi
-\frac{1}{2}\,t_{ij}\,\left(\delta_{\alpha\kappa}\,n_{ij}^\chi+
\delta_{\alpha\chi}\,n_{ij}^\kappa\right)
\quad.
\end{equation}
Inserting this expression in the sum~(\ref{eqn:xii:xiij}), the second term disappears
because the system is at mechanical equilibrium. This yields:
\begin{equation}
\label{eqn:xi:micro:pair}
{\Xi}^\alpha_{i,\kappa\chi} = -\sum_j\,
\left(r_{ij}\,c_{ij}-t_{ij}\right) n_{ij}^\alpha\,n_{ij}^\kappa\,n_{ij}^\chi
\quad.
\end{equation}

It is now an easy task to show 
that the terms in equation~(\ref{eqn:born:micro:xi}) involving contributions from the stress tensor disappear, 
which leads to the expression:
\begin{equation}
\label{eqn:born:micro:pair}
C^{\rm Born}_{\alpha\beta\kappa\chi}=\frac{1}{\mathring{V}}\,\sum_{ij}\left(r_{ij}\,c_{ij}-t_{ij}\right)
\,r_{ij}\,n_{ij}^\alpha\,n_{ij}^\beta\,n_{ij}^\kappa\,n_{ij}^\chi
\end{equation}

\end{document}

%% file: transformationCartoon.pstex_t
\begin{picture}(0,0)%
\includegraphics{transformationCartoon.pstex}%
\end{picture}%
\setlength{\unitlength}{4144sp}%
\begingroup\makeatletter\ifx\SetFigFont\undefined%
\gdef\SetFigFont#1#2#3#4#5{%
  \reset@font\fontsize{#1}{#2pt}%
  \fontfamily{#3}\fontseries{#4}\fontshape{#5}%
  \selectfont}%
\fi\endgroup%
\begin{picture}(4756,2205)(664,-4279)
\put(2476,-2581){\makebox(0,0)[lb]{\smash{{\SetFigFont{20}{24.0}{\familydefault}{\mddefault}{\updefault}{\color[rgb]{0,0,0}$\tensor {h}=\tensor{\mathring{h}}^{\prime}$}%
}}}}
\put(1036,-2581){\makebox(0,0)[lb]{\smash{{\SetFigFont{20}{24.0}{\familydefault}{\mddefault}{\updefault}{\color[rgb]{0,0,0}$\tensor{\mathring{h}}$}%
}}}}
\put(4501,-2581){\makebox(0,0)[lb]{\smash{{\SetFigFont{20}{24.0}{\familydefault}{\mddefault}{\updefault}{\color[rgb]{0,0,0}$\tensor h^{\prime}$}%
}}}}
\put(3331,-3481){\makebox(0,0)[lb]{\smash{{\SetFigFont{20}{24.0}{\rmdefault}{\mddefault}{\updefault}{\color[rgb]{0,0,0}$\tensor{F}^{\prime\prime}$}%
}}}}
\put(1531,-3481){\makebox(0,0)[lb]{\smash{{\SetFigFont{20}{24.0}{\rmdefault}{\mddefault}{\updefault}{\color[rgb]{0,0,0}$\tensor{F}$}%
}}}}
\put(2521,-4156){\makebox(0,0)[lb]{\smash{{\SetFigFont{20}{24.0}{\rmdefault}{\mddefault}{\updefault}{\color[rgb]{0,0,0}$\tensor{F}^{\prime}$}%
}}}}
\end{picture}%